\documentclass[pmlr]{jmlr}

\usepackage{array}
\usepackage{float}
\usepackage[font=small]{caption}
\usepackage{subcaption}
\usepackage{booktabs}
\usepackage{url}
\usepackage{amsmath}
\usepackage{amssymb}
\usepackage{algorithm2e}
\usepackage{physics}
\usepackage{stmaryrd}
\usepackage{graphicx}
\usepackage{enumitem}
\usepackage{scalerel}
\usepackage{xparse}
\usepackage{tikz}

\DeclareMathOperator\doo{do}

\DeclareMathOperator\an{an}
\DeclareMathOperator\pa{pa}
\DeclareMathOperator\de{de}

\DeclareMathOperator\nde{nde}

\DeclareMathOperator\dom{dom}

\newcommand{\M}{\mathcal{M}}
\newcommand{\Gm}{\mathcal{G}}
\newcommand{\U}{\mathcal{U}}
\newcommand{\V}{\mathcal{V}}

\newcommand{\E}{\mathbb{E}}

\providecommand{\fatsemi}{\,;\,}

\NewDocumentEnvironment{definition*}{o}
{%
  \par\medskip
  \noindent\textbf{Definition%
  \IfValueT{#1}{ (#1)}.}\ %
  \ignorespaces
}
{%
  \par\medskip
}
\NewDocumentEnvironment{example*}{o}
{%
  \par\medskip
  \noindent\textbf{Example%
  \IfValueT{#1}{ (#1)}.}\ %
  \ignorespaces
}
{%
  \par\medskip
}

\DeclareSymbolFont{rmlargesymbols}{OMX}{mdbch}{m}{n}
\DeclareMathSymbol{\rmintop}{\mathop}{rmlargesymbols}{82}
\newcommand\rmint{\scaleobj{0.9}{\rmintop\nolimits}}

\newcommand{\indep}{\perp \!\!\!\! \perp}
\newcommand{\given}{\,|\,}

\usetikzlibrary{arrows.meta}

\jmlrvolume{343}
\jmlryear{2026}
\jmlrworkshop{Probabilistic Graphical Models (PGM)}

\title[Scope-Restricted Backtracking Counterfactuals]{Scope-Restricted Backtracking Counterfactuals}

\author{
  \Name{Johan {de Aguas}} \Email{johanmd@uio.no}\\
  \addr Integreat: Norwegian Centre for Knowledge-driven Machine Learning, University of Oslo, Norway
  \AND
  \Name{Martin Jullum} \Email{jullum@nr.no}\\
  \addr Norwegian Computing Center, Norway
}
\editors{Gustau Camps-Valls, Manuele Leonelli and Gherardo Varando}

\begin{document}

\maketitle

\begin{abstract}
Interventional counterfactuals evaluate statements such as ``had $A$ been $a$'' by replacing the assignment for $A$ while keeping the unit's exogenous background fixed. Backtracking counterfactuals reverse this logic: they preserve the structural assignments and instead vary background conditions across worlds. This semantics is natural for diagnostic explanation, but full backtracking may be too permissive because every exogenous coordinate can vary. We introduce \emph{scope-restricted backtracking counterfactuals}, in which the modeler specifies which coordinates may vary while the remainder are shared across worlds. The scope determines the admissible carriers of the contrast, while a cross-world kernel governs the plausibility of their alternative values. We present model-based inference for general scope-restricted backtracking counterfactuals, derive sharp marginal bounds for a specified explanatory query, and establish frontier and compatibility conditions under which this explanatory query admits a modified-value policy representation. A linear-Gaussian example distinguishes local-noise, ancestor, irrelevant-noise, and infeasible explanations across scopes, while an application to the UCI Adult benchmark illustrates kernel specification, sensitivity, and how much of the explanation is carried by the cross-world coupling itself.
\end{abstract}


\section{Introduction}\label{sec:intro}

Counterfactual reasoning is central to explainable artificial intelligence (XAI), where explanations often ask what would have needed to be different for an observed outcome to change \citep{wachter2017counterfactual}. In \emph{algorithmic recourse}, this question is action-oriented: what feasible action could an agent take to obtain a more favorable outcome? \citep{ustun2019actionable,karimi2020algorithmic}. In \emph{diagnostic explanations}, by contrast, the question is contrastive but not necessarily actionable: what circumstances would have had to differ for an alternative outcome to occur? \citep{miller2019explanation,beckers2022causal,budhathoki2022causal}.

From a structural-causal perspective, counterfactuals are model-defined cross-world quantities rather than merely alternative feature vectors produced by an explanation method. Within this framework, action-oriented and diagnostic interpretations correspond naturally to interventional and backtracking semantics, respectively. Under an \emph{interventional counterfactual} \citep{PearlCausality}, the antecedent ``had $A$ been $a$'' is evaluated by replacing the structural assignment for $A$ while keeping the unit's background characteristics fixed. Under a \emph{backtracking counterfactual} \citep{vonKugelgen2023backtracking}, the structural assignments are left unchanged and the antecedent is instead realized by revising the exogenous background through a cross-world kernel. Counterfactual worlds thus remain on the structural manifold, closely connecting backtracking to the plausibility and on-manifold constraints used in XAI \citep{poyiadzi2020face,crupi2022counterfactual,Redelmeier2024MCCE}.

Much of practical XAI, however, generates ``counterfactuals'' without an explicit causal graphical model. The method of \citet{wachter2017counterfactual}, along with many approaches surveyed by \citet{verma2024counterfactual} and \citet{Guidotti2024}, searches directly in feature space for perturbations that alter a classifier's prediction. Other methods constrain this search to actionable or causally plausible regions: \citet{mothilal2020dice}, for example, allow features to be designated as immutable or bounded, while \citet{mahajan2019preserving} use partial structural models to preserve causal relations. Although such constraints can encode actionability or plausibility, without an explicit causal model they do not specify how changes to one feature propagate to its descendants.

Full backtracking uses structural information, but may be too permissive for explanation. In its unrestricted form, any exogenous coordinate may vary, including coordinates irrelevant to the antecedent or whose revision changes features intended to remain fixed. This conflicts with several common XAI goals: sparse explanations seek a small locus of change; recourse typically excludes immutable features such as age or place of birth; diagnostic queries may hold selected features fixed; and fairness queries may require fixing a protected attribute, varying it while preserving selected background facts, or restricting the pathways through which it affects a prediction \citep{kusner2017counterfactual,kilbertus2017avoiding,vonKugelgen2022fairness}. Because backtracking scope is defined over exogenous coordinates, translating such feature-level requirements into scope remains model-dependent.

This paper develops \emph{scope-restricted} backtracking counterfactuals. The modeler selects which exogenous coordinates may backtrack, while the remainder stay at their factual values. Restricted backtracking is suggested in \citet{vonKugelgen2023backtracking} and appears in related computational forms: \citet{kladny2024deep} impose sparsity on counterfactual representations, \citet{hao2024natural} restrict variation to a learned manifold, and \citet{fatemi2025brace} operationalize restricted backtracking with a deep generative model. Our focus is complementary: we give scope an explicit cross-world semantics, separate it from kernel specification, and characterize when the resulting diagnostic query admits a modified-value-policy representation. This gives the explanation a policy interpretation and, under additional actionability and effectiveness requirements, connects it to algorithmic recourse.

\paragraph{Contributions.}
We define scope-restricted backtracking through scoped cross-world laws, establish a simultaneous $\sigma$-regime representation, and develop model-based inference clarifying how scope, latent exogenous states, and kernel specification determine support-based antecedent feasibility. We show that if a downstream frontier intercepts all relevant effects of the scoped revision and query--response compatibility holds, the conditional explanation law admits a modified-value policy representation. When the full cross-world law is unavailable, Appendix~\ref{app:partial-identification} derives sharp marginal bounds for discrete or event-valued explanations from selected factual and counterfactual margins, together with denominator-based refinements. Finally, we illustrate scoped diagnostic explanations in XAI, their sensitivity to scope and kernel choices, and the additional requirements needed for a recourse interpretation.

\paragraph{Preliminaries.}
We work throughout with structural causal models (SCMs) by \citet{PearlCausality} and the associated semantics of hard and soft interventions by \citet{Correa_Bareinboim_2020,NeuripsCorrea}. Appendix~\ref{app:preliminaries} collects the formal definitions, notation, and theoretical background used in the paper, including the SCM solution map, graphical notation, hard-intervention counterfactuals, and the $\sigma$-regime formulation of soft interventions and its special cases.

Throughout, $\M=(\V,\U,\Gm,\mathcal F,P_{\U})$ denotes an SCM with endogenous variables $\V$, exogenous background $\U$, directed acyclic graph (DAG) $\Gm$, causal assignments $\mathcal F=\{f_V\}_{V\in\V}$, and background law $P_{\U}$; $U_V$ denotes the exogenous coordinates entering $f_V$, $\Upsilon_{\M}$ the SCM solution map, and $\pa(\cdot;\Gm)$, $\an(\cdot;\Gm)$, $\de(\cdot;\Gm)$, and $\nde(\cdot;\Gm)$ the usual graphical parent, ancestor, descendant, and non-descendant operators. Formal definitions are given in Appendix~\ref{app:preliminaries}.

With an abuse of notation, we use $\dd P$ both for the density/mass of $P$ with respect to a suitable dominating measure $\mu_{\mathrm{dom}}$ and for the corresponding probability element.

\section{Backtracking Counterfactuals}\label{sec:backtracking}

Backtracking counterfactuals \citep{vonKugelgen2023backtracking} keep the structural assignments of an SCM fixed and allow the exogenous background to differ across factual and counterfactual worlds. Instead of replacing components of $\mathcal F$, they augment $\M$ with a counterfactual copy of $\U$ and a cross-world \emph{backtracking conditional kernel} $B(u^\ast\given u)$. This kernel specifies which counterfactual backgrounds $u^\ast$ are considered plausible relative to a factual background $u$, before conditioning on evidence. It induces cross-world laws of the form
\begin{equation}\label{eq:back}
    \mathbb{P}(Y^{\ast}\in\mathcal{Y},Z\in\mathcal{Z})
    :=
    \rmint\!\!\!\rmint\!
    \mathbb{I}\{Y(u^\ast)\in \mathcal{Y}\}\,
    \mathbb{I}\{Z(u)\in \mathcal{Z}\}\,
    \dd B(u^\ast\given u)\,
    \dd P_{\U}(u),
\end{equation}
for measurable $\mathcal Y\subseteq\dom Y$ and $\mathcal Z\subseteq\dom Z$. Here $Y(u^\ast)$ is obtained by solving the unchanged structural equations $\Upsilon_{\mathcal{M}}$ at the counterfactual exogenous input $u^\ast$. Given factual evidence $H=h$, the factual background distribution $P_{\U}$ in \eqref{eq:back} is updated by abduction. Conditional backtracking queries are then obtained by ordinary probabilistic conditioning on the relevant factual and counterfactual events, whenever these have positive probability.

A common modeling choice is a distance-based Gibbs kernel $\dd B(u^\ast\given u)\propto \exp\{-\beta\,d(u^\ast,u)\}$, which favors counterfactual backgrounds close to the factual one. The choice of $B$ constitutes a cross-world structure: observational data may inform the distribution of factual states, but do not generally determine how a factual background $u$ should be paired with a counterfactual one $u^\ast$. A generative model may provide a latent representation and meaningful distances, but the cross-world coupling still requires a modeling choice, calibration, and sensitivity analysis. Under scope restriction, this distinction becomes explicit: the scope determines \emph{which} coordinates may change, while the kernel determines \emph{how} they change.

Two points are important for what follows. First, an antecedent may be infeasible: if $X(u^\ast)=x^\ast$ has no support under the unchanged structural assignments and the chosen kernel, the corresponding conditional law is undefined. Second, because full backtracking absorbs the contrast through the exogenous state, it may revise ancestors, local noises, and descendants of the antecedent, including coordinates unrelated to realizing it. Section~\ref{sec:example} illustrates both phenomena in a linear-Gaussian SCM.

\section{Restricted Backtracking Counterfactuals}\label{sec:partial}

Instead of drawing a wholly new background $u^\ast$, choose a scope $\mathcal E\subseteq\U$ of exogenous coordinates that may backtrack and keep $\bar{\mathcal E}=\U\setminus\mathcal E$ fixed. Write $u=(e,\bar e)$ and $u^\ast=(e^\ast,\bar e^\ast)$ for the corresponding decomposition into scoped and pinned coordinates.

\begin{definition*}[Scope-restricted backtracking kernel]\label{def:partial-kernel}
A \emph{scope-restricted backtracking kernel with scope $\mathcal E$} is a cross-world kernel of the form
\begin{equation}\label{eq:partial-kernel}
\dd B(u^\ast\given u)
=
\dd K(e^\ast\given u)\,\delta_{\bar e}(\bar e^\ast),
\end{equation}
where $K$ specifies how the coordinates in $\mathcal E$ may change and $\delta_{\bar e}$ denotes the density/mass of the Dirac measure at $\bar e$, which pins all coordinates in $\bar{\mathcal E}$ to their factual values. With a slight abuse of notation, \eqref{eq:partial-kernel} should be understood, at the level of distribution functions as $B(e^\ast,\bar e^\ast\given u) = K(e^\ast\given u)\mathbb I\{\bar e\leq\bar e^\ast\}$, with the inequality taken componentwise.
\end{definition*}


Given a factual background $u=(e,\bar e)$, draw $e^\ast\sim K(\cdot\given e,\bar e)$ and set $u^\ast=(e^\ast,\bar e)$. For any endogenous variable $V$, write $V^\ast:=V(e^\ast,\bar e)$ for its value obtained by solving the unchanged SCM at this partially revised background. Thus, for endogenous $Y,Z\subseteq\V$ and measurable $\mathcal Y,\mathcal Z$, the scoped cross-world law is
\begin{equation}\label{eq:partialback}
\mathbb{P}(Y^\ast\in\mathcal Y,Z\in\mathcal Z)
=
\rmint\!\!\!\rmint
\mathbb I\{Y(e^\ast,\bar e)\in\mathcal Y\}\,
\mathbb I\{Z(e,\bar e)\in\mathcal Z\}\,
\dd K(e^\ast\given e,\bar e)\,
\dd P_{\U}(e,\bar e).
\end{equation}
If $\mathcal E=\varnothing$, this gives the observational law; if $\mathcal E=\U$ and $K=B$, it gives full backtrack \eqref{eq:back}.

Scope restriction is not the same as querying a factual variable alongside a counterfactual one. If $V$ is downstream of the antecedent, the factual value $V(u)$ remains unchanged and therefore does not propagate the counterfactual contrast. By contrast, the full backtracking value $V(u^\ast)$ propagates the contrast while also allowing its local noise $u_V^\ast$ to change, whereas the restricted value $V(e^\ast,\bar e)$ propagates the contrast while retaining the factual noise whenever $U_V\in\bar{\mathcal E}$. Thus, restricted backtracking is a sub-family of full backtracking obtained through kernel \eqref{eq:partial-kernel}, but it makes non-revision of $\bar{\mathcal E}$ a hard constraint rather than merely encouraging it through a distance penalty. Example~\ref{ex:linear} shows the consequence: pinning $U_Z$ removes the irrelevant $\sigma_Z^2$ variance contributed by full backtracking.

\begin{figure*}[t]
\centering

\tikzset{
  var/.style={inner sep=1pt, font=\small},
  exo/.style={inner sep=1pt, font=\small},
  int/.style={draw, rectangle, inner sep=2pt, font=\small},
  arr/.style={->, >=Stealth, line width=1.0pt},
  back/.style={->, >=Stealth, densely dashed, line width=1.0pt}
}

\begin{minipage}[t]{0.235\textwidth}
\centering
\resizebox{\linewidth}{!}{%
\begin{tikzpicture}

\node[var] (X)  at (0,1) {$X$};
\node[var] (Y)  at (0,0) {$A$};
\node[var] (Z)  at (0,-1) {$Z$};

\node[var] (Xs) at (3.1,1) {$X$};
\node[int] (Ys) at (3.1,0) {$a$};
\node[var] (Zs) at (3.1,-1) {$Z^a$};

\node[exo,red] (Ux) at (1.55,1.75) {$U_X$};
\node[exo,red] (Uy) at (1.55,0.55) {$U_A$};
\node[exo,red] (Uz) at (1.55,-0.55) {$U_Z$};

\draw[arr,gray] (Ux) -- (X);
\draw[arr,gray] (Uy) -- (Y);
\draw[arr,gray] (Uz) -- (Z);
\draw[arr] (X) -- (Y);
\draw[arr] (Y) -- (Z);
\draw[arr] (X) to[bend right=35] (Z);

\draw[arr,gray] (Ux) -- (Xs);
\draw[arr,gray] (Uz) -- (Zs);
\draw[arr] (Ys) -- (Zs);
\draw[arr] (Xs) to[bend left=35] (Zs);

\end{tikzpicture}%
}

\vspace{1mm}
{\small (a) Interventional}
\end{minipage}
\hfill
\begin{minipage}[t]{0.235\textwidth}
\centering
\resizebox{\linewidth}{!}{%
\begin{tikzpicture}

\node[var] (X)  at (0,1) {$X$};
\node[var] (Y)  at (0,0) {$A$};
\node[var] (Z)  at (0,-1) {$Z$};

\node[var] (Xs) at (3.3,1) {$X^\ast$};
\node[var] (Ys) at (3.3,0) {$A^\ast$};
\node[var] (Zs) at (3.3,-1) {$Z^\ast$};

\node[exo,red] (Ux)  at (1.05,1.75) {$U_X$};
\node[exo,red] (Uy)  at (1.05,0.55) {$U_A$};
\node[exo,red] (Uz)  at (1.05,-0.55) {$U_Z$};

\node[exo,red] (Uxs) at (2.25,1.75) {$U_X^\ast$};
\node[exo,red] (Uys) at (2.25,0.55) {$U_A^\ast$};
\node[exo,red] (Uzs) at (2.25,-0.55) {$U_Z^\ast$};

\draw[back,blue] (Ux) -- (Uxs);
\draw[back,blue] (Uy) -- (Uys);
\draw[back,blue] (Uz) -- (Uzs);

\draw[arr,gray] (Ux) -- (X);
\draw[arr,gray] (Uy) -- (Y);
\draw[arr,gray] (Uz) -- (Z);
\draw[arr] (X) -- (Y);
\draw[arr] (Y) -- (Z);
\draw[arr] (X) to[bend right=35] (Z);

\draw[arr,gray] (Uxs) -- (Xs);
\draw[arr,gray] (Uys) -- (Ys);
\draw[arr,gray] (Uzs) -- (Zs);
\draw[arr] (Xs) -- (Ys);
\draw[arr] (Ys) -- (Zs);
\draw[arr] (Xs) to[bend left=35] (Zs);

\end{tikzpicture}%
}

\vspace{1mm}
{\small (b) Full backtracking}
\end{minipage}
\hfill
\begin{minipage}[t]{0.235\textwidth}
\centering
\resizebox{\linewidth}{!}{%
\begin{tikzpicture}

\node[var] (X)  at (0,1) {$X$};
\node[var] (Y)  at (0,0) {$A$};
\node[var] (Z)  at (0,-1) {$Z$};

\node[var] (Xs) at (3.3,1) {$X$};
\node[var] (Ys) at (3.3,0) {$A^\ast$};
\node[var] (Zs) at (3.3,-1) {$Z^\ast$};

\node[exo,red] (Ux) at (1.65,1.75) {$U_X$};
\node[exo,red] (Uz) at (1.65,-0.55) {$U_Z$};

\node[exo,red] (Uy)  at (1.05,0.55) {$U_A$};
\node[exo,red] (Uys) at (2.25,0.55) {$U_A^\ast$};

\draw[back,blue] (Uy) -- (Uys);

\draw[arr,gray] (Ux) -- (X);
\draw[arr,gray] (Uy) -- (Y);
\draw[arr,gray] (Uz) -- (Z);
\draw[arr] (X) -- (Y);
\draw[arr] (Y) -- (Z);
\draw[arr] (X) to[bend right=35] (Z);

\draw[arr,gray] (Ux) -- (Xs);
\draw[arr,gray] (Uys) -- (Ys);
\draw[arr,gray] (Uz) -- (Zs);
\draw[arr] (Xs) -- (Ys);
\draw[arr] (Ys) -- (Zs);
\draw[arr] (Xs) to[bend left=35] (Zs);

\end{tikzpicture}%
}

\vspace{1mm}
{\small (c) $\mathcal E=\{U_A\}$}
\end{minipage}
\hfill
\begin{minipage}[t]{0.235\textwidth}
\centering
\resizebox{\linewidth}{!}{%
\begin{tikzpicture}

\node[var] (X)  at (0,1) {$X$};
\node[var] (Y)  at (0,0) {$A$};
\node[var] (Z)  at (0,-1) {$Z$};

\node[var] (Xs) at (3.3,1) {$X^\ast$};
\node[var] (Ys) at (3.3,0) {$A^\ast$};
\node[var] (Zs) at (3.3,-1) {$Z^\ast$};

\node[exo,red] (Ux)  at (1.05,1.75) {$U_X$};
\node[exo,red] (Uy)  at (1.05,0.55) {$U_A$};

\node[exo,red] (Uxs) at (2.25,1.75) {$U_X^\ast$};
\node[exo,red] (Uys) at (2.25,0.55) {$U_A^\ast$};

\node[exo,red] (Uz) at (1.65,-0.55) {$U_Z$};

\draw[back,blue] (Ux) -- (Uxs);
\draw[back,blue] (Uy) -- (Uys);

\draw[arr,gray] (Ux) -- (X);
\draw[arr,gray] (Uy) -- (Y);
\draw[arr,gray] (Uz) -- (Z);
\draw[arr] (X) -- (Y);
\draw[arr] (Y) -- (Z);
\draw[arr] (X) to[bend right=35] (Z);

\draw[arr,gray] (Uxs) -- (Xs);
\draw[arr,gray] (Uys) -- (Ys);
\draw[arr,gray] (Uz) -- (Zs);
\draw[arr] (Xs) -- (Ys);
\draw[arr] (Ys) -- (Zs);
\draw[arr] (Xs) to[bend left=35] (Zs);

\end{tikzpicture}%
}

\vspace{1mm}
{\small (d) $\mathcal E=\{U_X,U_A\}$}
\end{minipage}

\caption{
Twin-network \citep{BalkeTwin} comparison of interventional, full-backtracking, and scope-restricted backtracking in a Markovian SCM.
(a) The antecedent $A$ is fixed by an intervention to $a$, represented by the boxed node, and its incoming arrows are removed.
(b) Under full backtracking, all exogenous coordinates may differ across factual and counterfactual worlds, coupled by $B$.
(c) Under restricted backtracking with scope $\mathcal E=\{U_A\}$, only the local exogenous input of $A$ may change, while all remaining ones are shared across worlds.
(d) Under scope $\mathcal E=\{U_X,U_A\}$, the background of the ancestor $X$ may also change, while $U_Z$ remains pinned to its factual value.
}
\label{fig:partial-backtracking-four-panels}
\end{figure*}
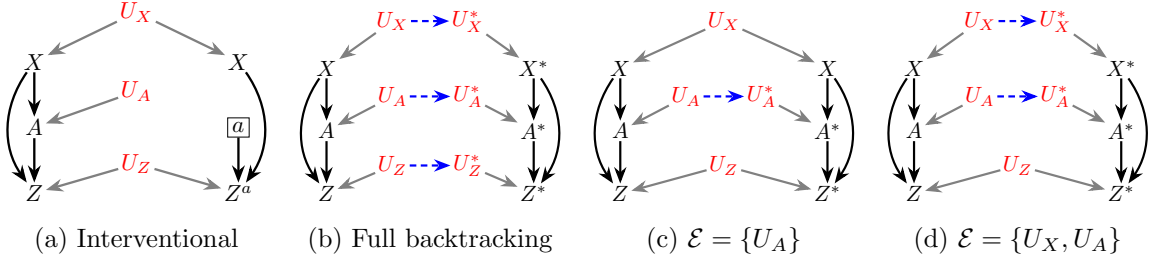

\subsection{Choosing the scope and kernel}\label{sec:kernel-practice}
 
The scope and kernel encode distinct modeling choices. The scope $\mathcal E$ determines which exogenous coordinates are permitted to differ across worlds, while $K$ determines which revisions of those coordinates are supported and how they are weighted. Thus, changing the scope alters the admissible directions of revision, whereas, for a fixed support, changing $K$ changes their relative plausibility. We say that $K$ is \emph{decomposable} if, for every $u$, one has $K(e^\ast\given u) = \prod_{U_j\in\mathcal E}K_j(u_j^\ast\given u_j)$. 

Figure~\ref{fig:partial-backtracking-four-panels} contrasts interventional and full-backtracking counterfactuals with two scope-restricted constructions under a decomposable kernel.

There are several practical ways to specify $K$. One option is to impose a structurally motivated \emph{close-to-data-manifold} kernel, for example $\dd K(e^\ast\given u)\propto \exp\{-\beta\,d_{\mathcal E}(e^\ast,e)\}$ or, in a decomposable continuous model, $U_j^\ast\given (U_j\!=\!u_j)\sim \mathcal{N}(u_j,\kappa_j^2)$ for $U_j\in\mathcal{E}$. Here, $d_{\mathcal E}$ encodes which revisions are considered similar, while $\beta$ or $\kappa_j$ controls the strength of cross-world similarity. Alternatively, a fitted generative model may provide latent coordinates and a useful geometry for defining $d_{\mathcal E}$. When no single kernel is clearly defensible, $\beta$, $\kappa_j$, or other aspects of $K$ can instead be treated as sensitivity parameters, or one may target the partial-identification analysis in Appendix~\ref{app:partial-identification}.


Scope restriction provides control over which background coordinates may change, but an overly narrow scope may prevent the desired counterfactual contrast from being realized at all. Feasibility is therefore a prerequisite for a conditional backtracking query and provides a basic criterion for scope selection. Let $V\subseteq\V$ denote the endogenous variables appearing in the antecedent. Given factual evidence $Z=z$, we say that an antecedent $V^\ast=v$ is \emph{feasible} under scope $\mathcal E$ if $v$ lies in the support of the scoped conditional law of $V^\ast$ given $Z=z$. In the discrete case, this reduces to $\mathbb P(V^\ast=v,Z=z)>0$ whenever the factual event $Z=z$ itself has positive probability. This yields a simple monotonicity principle for nested scopes.

\begin{proposition}[Support monotonicity]\label{prop:support-monotonicity}
Let $\mathcal E_1\subseteq\mathcal E_2$, write $\mathcal D=\mathcal E_2\setminus\mathcal E_1$, and let $K_{\mathcal E_j}(\cdot\given u)$ denote the scope kernel associated with $\mathcal E_j$, $j\in\{1,2\}$. Suppose that, for every factual background $u$, $\operatorname{supp}\{K_{\mathcal E_1}(\cdot\given u)\}\times\{d(u)\}
\subseteq
\operatorname{supp}\{K_{\mathcal E_2}(\cdot\given u)\}$, where $d(u)$ denotes the factual value of the coordinates in $\mathcal D$. Then every antecedent feasible under $\mathcal E_1$ is feasible under $\mathcal E_2$.
\end{proposition}

The property does not compare scopes that are not nested, since different scopes may activate distinct paths for realizing the same antecedent (see Appendix~\ref{app:support-monotonicity} and Example~\ref{ex:linear}).

\subsection{Model-based counterfactual inference}\label{sec:properties}

Suppose the relevant structural mechanisms of the SCM are known, or are represented by a fitted generative model that supports abduction and prediction, and let the scope-restricted kernel $K$ be specified. Conditional queries such as $\mathbb{P}(X^\ast\in\mathcal X\given Y^\ast=y^\ast,Z=z)$, where $Z$ here denotes generically the available factual evidence, can then be computed by the usual abduction--prediction logic, subject to the restriction $\bar e^\ast=\bar e$:
\begin{enumerate}[leftmargin=5mm]\setlength\itemsep{0em}
\item \textbf{Partial cross-world abduction.}
Update the joint law $K(e^\ast\given e,\bar e)\,P_{\U}(e,\bar e)$ using the counterfactual and factual evidence:
\begin{equation*}
\dd P_{\mathrm{post}}(e^\ast,e,\bar e\given Y^\ast=y^\ast,Z=z)
\propto
\mathbb I\{Y(e^\ast,\bar e)=y^\ast\}\,
\mathbb I\{Z(e,\bar e)=z\}\,
\dd K(e^\ast\given e,\bar e)\,
\dd P_{\U}(e,\bar e),
\end{equation*} 
with normalizing constant obtained by integration over $(e^\ast,e,\bar e)$.

\item \textbf{Marginalization.}
Remove factual coordinates not needed for prediction:
\begin{equation*}
\dd P_{\mathrm{marg}}(u^\ast\given Y^\ast=y^\ast,Z=z)
=
\rmint\!
\dd P_{\mathrm{post}}(e^\ast,e,\bar e\given Y^\ast=y^\ast,Z=z)\, \dd\mu_{\mathrm{dom}}(e),
\end{equation*}
where $u^\ast=(e^\ast,\bar e)$. Thus, the resulting posterior is supported only on counterfactual worlds sharing the factual coordinates $\bar e$.

\item \textbf{Prediction.}
Evaluate the unchanged structural assignments under this posterior:
\begin{equation*}
\mathbb{P}(X^\ast\in\mathcal X\given Y^\ast=y^\ast,Z=z)
=
\rmint\!
\mathbb I\{X(u^\ast)\in\mathcal X\}\,
\dd P_{\mathrm{marg}}(u^\ast\given Y^\ast=y^\ast,Z=z).
\end{equation*}
\end{enumerate}
For continuous conditioning variables, the displays above are understood through the corresponding densities or regular conditional distributions.

\subsection{Scoped counterfactual explanations}\label{sec:explanations}

A natural counterfactual-explanation query is to rank alternative values of endogenous explanatory variables $X\subseteq\nde(Y;\Gm)$ that accompany a desired outcome $Y^\ast=y^\ast$ under the chosen scope, conditional on the factual information available for the unit.

\begin{definition*}[MAP scoped backtracking explanation]\label{def:map-explanation}
Let $H=(X,Y,Z)$ collect the factual evidence, with realization $h=(x,y,z)$, where $Z=z$ denotes any additional pre-$X$ factual information beyond the explanatory variables $X=x$ and factual outcome $Y=y$. Under scope $\mathcal E\subseteq\U$, the posterior predictive density or mass
\begin{equation}\label{eq:query}
p_{X^\ast}(x^\ast\given Y^\ast=y^\ast,H=h)
\end{equation}
ranks candidate values $x^\ast$ among scoped counterfactual worlds consistent with the factual evidence and realizing the desired outcome \citep{fatemi2025brace}. Here, $X^\ast$ is induced by revisions of the exogenous coordinates in $\mathcal E$ propagated through the unchanged structural assignments, so $\mathcal E$ need not correspond directly to the coordinates of $X$. A \emph{maximum a posteriori} (MAP) scoped backtracking explanation is any
\begin{equation}\label{eq:queryMAP}
x^\ast_{\mathrm{MAP}}
\in
\arg\max_{x^\ast}
p_{X^\ast}(x^\ast\given Y^\ast=y^\ast,H=h).
\end{equation}
\end{definition*}

This posterior-mode view parallels mode-based backtracking explanations \citep{kladny2024deep,kinjo2026generalized}, while related probabilistic approaches characterize explanations through the full posterior \citep{raman2023bayesian}. Intuitively, $x^\ast_{\mathrm{MAP}}$ is the most plausible value of $X$ among scoped worlds that agree with the factual evidence $H=h$ and realize the desired contrast $Y^\ast=y^\ast$. It thus provides a point-valued diagnostic explanation, with admissible background revisions controlled by $\mathcal E$ and their relative plausibility governed by $K$.

\paragraph{Partial identification.}
When a fitted generative model for the SCM is unavailable, or the scoped cross-world law is only partially specified, the explanation query in \eqref{eq:query} may not be point identified. Proposition~\ref{prop:xai-pearl-bounds} in Appendix~\ref{app:partial-identification} derives sharp marginal bounds for discrete or event-valued versions of this target from selected factual and counterfactual margins. The appendix also gives explicit refinements when the query denominator is known or bounded and a general feasible-set template for incorporating additional kernel restrictions.

\section{Backtracking Counterfactuals and Modified-Value Policy Frontiers}
\label{sec:soft-bridge}

Restricted backtracking acts by revising part of the exogenous background and then propagating the original structural assignments. This section asks when the same explanatory behavior can instead be represented more locally, by modifying a collection of \emph{endogenous} variables. Specifically, we ask whether the effect of the scoped revision on the pair $(X^\ast,Y^\ast)$ can be ``pushed forward'' to an endogenous \emph{frontier} $A\subseteq\V$: once the revised value $A^\ast$ at that frontier is known, can the upstream background revision be discarded and replaced by an intervention on $A$? 

Such a reduction is useful because it connects a backtracking query to the language of stochastic and modified-value interventions \citep{haneuse2013estimation,sani2020identification} without requiring the entire scoped cross-world mechanism to admit such a representation. The frontier need not coincide with the explanatory variable $X$: it may lie upstream or downstream of $X$, contain $X$ itself, or consist of several variables jointly. Moreover, the induced policy may depend on the natural frontier value $A^\natural=A(u)$. This dependence is important: a backtracking revision can modify a variable differently depending on where that variable started, and conditioning on $A^\natural$ retains this factual--counterfactual coupling.

A simultaneous $\sigma$-regime representation of the full scoped law is given in Appendix~\ref{app:sigma-representations}. Here the aim is deliberately more local: to reproduce only the joint behavior of $(X^\ast,Y^\ast)$ needed by the explanatory query~\eqref{eq:query}.

\subsection{Reducibility to a modified-value policy}
\label{sec:reduction}

Let
\begin{equation}\label{eq:affected-set}
S_{\mathcal E}
:=
\{V\in\V:U_V\cap\mathcal E\ne\varnothing\}
\end{equation}
denote the endogenous variables whose \emph{local} exogenous arguments can be directly altered by the scope. For a candidate frontier $A\subseteq\V$, let
\begin{equation*}
C:=(X,Y),\qquad C^a(u)=(X^a(u),Y^a(u)).
\end{equation*}
The latter is the value of this target pair under $\operatorname{do}(A=a)$ and factual background $\U=u$.

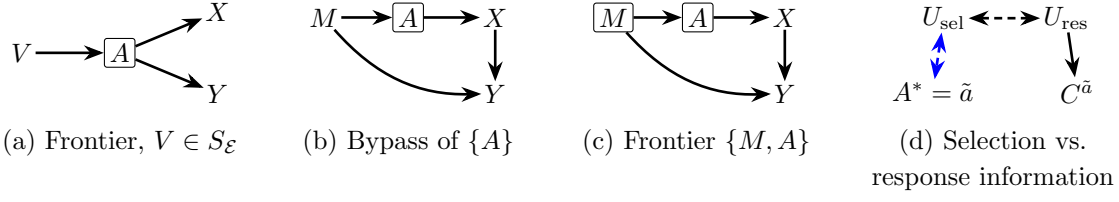
\begin{figure*}[t]
\centering

\tikzset{
  var/.style={inner sep=1pt, font=\small},
  exo/.style={inner sep=1pt, font=\small},
  int/.style={draw, rectangle, rounded corners=1pt, inner sep=2pt, font=\small},
  arr/.style={->, >=Stealth, line width=1.0pt},
  back/.style={<->, >=Stealth, densely dashed, line width=1.0pt}
}

\begin{minipage}[t]{0.235\textwidth}
\centering
\begin{tikzpicture}[scale=0.72]

\node[var] (V) at (0,0) {$V$};
\node[int] (A) at (1.8,0) {$A$};
\node[var] (X) at (3.6,0.75) {$X$};
\node[var] (Y) at (3.6,-0.75) {$Y$};

\draw[arr] (V) -- (A);
\draw[arr] (A) -- (X);
\draw[arr] (A) -- (Y);

\end{tikzpicture}

\vspace{1mm}
{\small (a) Frontier, $V\in S_{\mathcal E}$}
\end{minipage}
\hfill
\begin{minipage}[t]{0.235\textwidth}
\centering
\begin{tikzpicture}[scale=0.72]

\node[var] (M) at (0,0.65) {$M$};
\node[int] (A) at (1.55,0.65) {$A$};
\node[var] (X) at (3.15,0.65) {$X$};
\node[var] (Y) at (3.15,-0.75) {$Y$};

\draw[arr] (M) -- (A);
\draw[arr] (A) -- (X);
\draw[arr] (X) -- (Y);
\draw[arr] (M) to[bend right=25] (Y);

\end{tikzpicture}

\vspace{1mm}
{\small (b) Bypass of $\{A\}$}
\end{minipage}
\hfill
\begin{minipage}[t]{0.235\textwidth}
\centering
\begin{tikzpicture}[scale=0.72]

\node[int] (M) at (0,0.65) {$M$};
\node[int] (A) at (1.55,0.65) {$A$};
\node[var] (X) at (3.15,0.65) {$X$};
\node[var] (Y) at (3.15,-0.75) {$Y$};

\draw[arr] (M) -- (A);
\draw[arr] (A) -- (X);
\draw[arr] (X) -- (Y);
\draw[arr] (M) to[bend right=25] (Y);

\end{tikzpicture}

\vspace{1mm}
{\small (c) Frontier $\{M,A\}$}
\end{minipage}
\hfill
\begin{minipage}[t]{0.25\textwidth}
\centering
\begin{tikzpicture}[scale=0.72]

\node[exo] (Us) at (0.55,1.35) {$U_{\mathrm{sel}}$};
\node[exo] (Ur) at (2.80,1.35) {$U_{\mathrm{res}}$};
\node[var] (As) at (0.35,0) {$A^\ast=\tilde a$};
\node[var] (Ca) at (3.0,0) {$C^{\tilde a}$};

\draw[back,blue] (Us) -- (As);
\draw[arr] (Ur) -- (Ca);
\draw[back] (Us) -- (Ur);

\end{tikzpicture}

\vspace{1mm}
{\small (d) Selection vs. response information}
\end{minipage}

\caption{
Two requirements for reducibility. Boxes indicate the candidate frontier.
(a) If all relevant effects of the scoped revision pass through $A$, then fixing $A$ to its revised value can absorb the upstream change.
(b) With scope affecting $M$ and $A$, the route $M\to Y$ bypasses the singleton frontier $\{A\}$.
(c) Enlarging the frontier to $\{M,A\}$ removes this structural bypass.
(d) Structural interception is not sufficient: $A^\ast$ may depend on latent selection information $U_{\mathrm{sel}}$, while $C^{\tilde a}$ depends on latent response information $U_{\mathrm{res}}$. Conditional dependence between these components can make selection response-informative; Corollary~\ref{cor:graphical-frontier}(ii) rules this out through a latent-separation criterion. 
}
\label{fig:frontier-reducibility}
\end{figure*}

Two distinct obstacles can prevent such a reduction. The first is \emph{structural}: a scoped revision may reach $(X,Y)$ through a directed route that bypasses the proposed frontier. The second is \emph{probabilistic}: even after all relevant directed effects pass through $A$, the particular value $A^\ast$ selected by the backtracking construction may reveal latent factual information that changes how $(X,Y)$ respond to that value. Figure~\ref{fig:frontier-reducibility} illustrates these two obstacles.

\begin{proposition}[Reducibility through an endogenous frontier]
\label{prop:local-reducibility-mtp}
Let $\mathcal E\subseteq\U$ be a restricted backtracking scope, let $A\subseteq\V$ be a candidate endogenous frontier, and fix factual evidence $H=h$. Let $A^\natural:=A(u)$ denote the natural value of the frontier. Suppose:

\begin{enumerate}[label=(\roman*),leftmargin=6mm]\setlength\itemsep{0em}

\item \textbf{Downstream frontier.}
For every factual background $u=(e,\bar e)$ compatible with $H=h$ and every admissible scoped revision $e^\ast$, let $u^\ast=(e^\ast,\bar e)$. Then
\begin{equation}\label{eq:functional-frontier}
C(u^\ast)
=
C^{A(u^\ast)}(u).
\end{equation}
Thus, once the scoped revision has selected $A^\ast=A(u^\ast)$, assigning that same frontier value in the factual background reproduces both $X^\ast$ and $Y^\ast$.

\item \textbf{Query--response compatibility.}
Let $W\subseteq\nde(A;\Gm)$ denote observed pre-frontier information allowed to enter the policy. The variables $W$ and $H$ may overlap; whenever both appear, conditioning is understood on their joint information. For every $\tilde a$ in the conditional support of $A^\ast$,
\begin{equation}\label{eq:query--response-compatibility}
P_{C^{\tilde a}}
\bigl(
c\given
A^\ast=\tilde a,
W=w,
A^\natural=a^\natural,
H=h
\bigr)
=
P_{C^{\tilde a}}
\bigl(
c\given
W=w,
A^\natural=a^\natural,
H=h
\bigr)
\end{equation}
for every $c$. Thus, among factual units with the same $(W,A^\natural,H)$, learning that the backtracking construction selected $\tilde a$ provides no additional information about their response $C^{\tilde a}$ to that same frontier value; that is, selection itself is not response-informative.
\end{enumerate}

Define the modified-value policy
\begin{equation}\label{eq:q-natural-frontier}
\nu_h(\cdot\given w,a^\natural)
:=
P_{A^\ast}
\bigl(
\cdot\given
W=w,A^\natural=a^\natural,H=h
\bigr).
\end{equation}
After abduction on $H=h$, let the policy retain $(W,A^\natural)$, draw $A^{\nu_h} \sim \nu_h(\cdot\given W,A^\natural)$ independently of the remaining latent background conditional on $(W,A^\natural,H)$, and propagate the original structural assignments. Then
\begin{equation}\label{eq:query--reducibility}
p_{X^\ast}
(x^\ast\given Y^\ast=y^\ast,H=h)
=
p_{X^{\nu_h}}
(x^\ast\given Y^{\nu_h}=y^\ast,H=h).
\end{equation}
\end{proposition}

The proposition separates reducibility into a causal-path condition and a dependence condition. Condition~\emph{(i)} gives the unit-level identity $C^\ast=C^{A^\ast}$: once the scoped revision has selected its frontier value, its downstream effect on $(X,Y)$ can be reproduced by assigning that same value in the factual background. Condition~\emph{(ii)} then asks whether the remaining dependence between the selected value and the factual latent state can safely be discarded. A proof is given in Appendix~\ref{app:local-reducibility-proof}. Although conditions~\emph{(i)}--\emph{(ii)} are stated at the level needed for reducibility, they can be guaranteed by stronger but more readily interpretable graphical and structural criteria. The following corollary gives one such pair: a directed-cut criterion for the frontier and a latent-separation criterion ensuring query--response compatibility.

\begin{corollary}[Graphical and structural sufficient criteria]
\label{cor:graphical-frontier}
Suppose: \emph{(i)} the directly scope-affected target variables are themselves included in the frontier, $S_{\mathcal E}\cap(X\cup Y)\subseteq A$ and, for every $V\in S_{\mathcal E}\setminus A$ and $V^{\dagger}\in(X\cup Y)\setminus A$,
\begin{equation}\label{eq:graphical-frontier}
\text{every directed path }V\rightarrow\!\cdot\!\rightarrow V^{\dagger}
\text{ in }\Gm\text{ intersects }A;
\end{equation}
and \emph{(ii)} there exist latent subvectors
$U_{\mathrm{sel}},U_{\mathrm{res}}\subseteq\U$,
with complements in $\U$ denoted by
$\overline U_{\mathrm{sel}}$ and $\overline U_{\mathrm{res}}$.
Let $Q=(W,A^\natural)$ and $L=(Q,H)=(W,A^\natural,H)$. For every $\tilde a$ in the
conditional support of $A^\ast$, suppose
\begin{equation}\label{eq:structural-comp}
A^\ast
\indep \overline U_{\mathrm{sel}}
    \given U_{\mathrm{sel}},L,\quad
C^{\tilde a}
\indep \overline U_{\mathrm{res}}
    \given U_{\mathrm{res}},L,\quad
U_{\mathrm{sel}}
\indep U_{\mathrm{res}}
    \given L.
\end{equation}
Then condition \emph{(i)} implies the downstream-frontier
condition~\eqref{eq:functional-frontier}, while condition \emph{(ii)}
implies the query--response compatibility
condition~\eqref{eq:query--response-compatibility}. Consequently, the
scoped explanatory query~\eqref{eq:query} is reducible to the
modified-value query~\eqref{eq:query--reducibility}.
\end{corollary}

Condition \eqref{eq:graphical-frontier} requires the frontier to contain any target directly affected by the scope and to intercept every remaining directed route from a scope-affected mechanism to the target pair $(X,Y)$. Condition~\eqref{eq:structural-comp}, in turn, separates latent information relevant to selecting $A^\ast$ from latent information relevant to the responses $C^{\tilde a}$, conditional on $L=(W,A^\natural,H)$. Thus, the first criterion can be assessed from the causal graph together with the scope-to-mechanism map $S_{\mathcal E}$, whereas the second concerns the joint SCM--backtracking construction. Both are sufficient rather than necessary, but together they provide a concrete way to diagnose when the more general conditions of Proposition~\ref{prop:local-reducibility-mtp} are satisfied.

Overall, the reduction clarifies when a scoped backtracking explanation can be represented by a modified-value policy: the scoped change must be absorbed by an endogenous frontier, and the way that frontier value is selected must not carry additional response-relevant information. When either requirement fails, the backtracking query remains genuinely a cross-world counterfactual. When reduction does hold, the resulting policy should still be interpreted according to the information it uses: if it depends on post-outcome evidence such as $Y$, it is a diagnostic representation. A prospective deployment interpretation additionally requires the policy inputs to be available before frontier assignment; a recourse interpretation further requires the frontier variables to be manipulable or actionable and the induced policy to satisfy an appropriate effectiveness criterion.

\section{A Linear-Gaussian Example}\label{sec:example}

This example makes the roles of scope, feasibility, and kernel choice concrete. Different scopes isolate local-noise and ancestor explanations, exclude irrelevant downstream revisions, or make the antecedent infeasible.

\begin{example}\label{ex:linear}
Consider the Markovian SCM
\begin{equation}\label{eq:linear-scm}
    X:=U_X,
    \qquad
    Y:=X+U_Y,
    \qquad
    Z:=X+Y+U_Z,
\end{equation}
with factual observation $(X,Y,Z)=(1,2,2)$, so $u=(U_X,U_Y,U_Z)=(1,1,-1)$. Notice that the exogenous realization is not separately observed here; it is recovered by deterministic abduction from the structural assignments. We ask the diagnostic question: what would $(X^\ast,Z^\ast)$ have been, had $Y^\ast=3$? A hard intervention $\doo(Y=3)$ changes the mechanism for $Y$, whereas full backtracking keeps all mechanisms unchanged and allows all exogenous coordinates to differ. Scope restriction lies between these extremes by specifying which coordinates may carry the contrast. Table~\ref{tab:example-comparison} summarizes the resulting explanations.

For a Gaussian full-backtracking kernel
$B(u^\ast\given u)\equiv\mathcal{N}\{u,\operatorname{diag}(\sigma_X^2,\sigma_Y^2,\sigma_Z^2)\}(u^\ast)$,
which is decomposable with coordinate-wise conditional factors $\mathcal{N}(u_X,\sigma_X^2)$, $\mathcal{N}(u_Y,\sigma_Y^2)$, and $\mathcal{N}(u_Z,\sigma_Z^2)$,
conditioning on $Y^\ast=3$ imposes $U_Y^\ast=3-U_X^\ast$. Define
\begin{equation*}
    \xi
    :=
    \frac{\sigma_Y^2}{\sigma_Y^2+\sigma_X^2}\in(0,1).
\end{equation*}
Then
\begin{equation*}
    (X^\ast,Z^\ast)\given(Y^\ast=3,u)
    \sim
    \mathcal{N}\!\left(\!\!
    \begin{pmatrix}
    2-\xi\\
    4-\xi
    \end{pmatrix}\!,\xi\sigma_X^2
        \begin{pmatrix}
        1 & 1\\
        1 & 1
        \end{pmatrix}+\sigma_Z^2
        \begin{pmatrix}
        0 & 0\\
        0 & 1
        \end{pmatrix}\!\!
    \right)\!.
\end{equation*}
Full backtracking therefore allows $U_Z^\ast$ to vary even though $U_Z^\ast$ is not needed to realize the antecedent. Restricted scopes separate the possible explanations:
\begin{enumerate}[leftmargin=5mm]\setlength\itemsep{0em}
    \item $\mathcal{E}=\{U_Y\}$ pins $U_X^\ast=1$ and $U_Z^\ast=-1$; the antecedent forces $U_Y^\ast=2$, so $(X^\ast,Z^\ast)=(1,3)$. This agrees with the hard counterfactual after $\doo(Y=3)$.

    \item $\mathcal{E}=\{U_X\}$ pins $U_Y^\ast=1$ and $U_Z^\ast=-1$; the antecedent forces $U_X^\ast=2$, so $(X^\ast,Z^\ast)=(2,4)$. The contrast in $Y$ is explained by a different upstream $X$.

    \item $\mathcal{E}=\{U_X,U_Y\}$ with independent Gaussian coordinate kernels gives
    \begin{equation*}
        U_X^\ast\given(Y^\ast=3,u)
        \sim
        \mathcal{N}\!\left(2-\xi,\xi\sigma_X^2\right),
        \qquad
        U_Y^\ast=3-U_X^\ast,
    \end{equation*}
    and $U_Z^\ast=-1$ is pinned. Hence
    \begin{equation*}
        (X^\ast,Z^\ast)\given(Y^\ast=3,u)
        \sim
        \mathcal{N}\!\left(\!\!
        \begin{pmatrix}
        2-\xi\\
        4-\xi
        \end{pmatrix}\!,\xi\sigma_X^2
        \begin{pmatrix}
        1 & 1\\
        1 & 1
        \end{pmatrix}\!\!
        \right)\!.
    \end{equation*}
    The mean agrees with full backtracking, but the variance contribution $\sigma_Z^2$ disappears.

    \item $\mathcal{E}=\{U_Z\}$ makes the antecedent infeasible. Since $U_X^\ast$ and $U_Y^\ast$ are pinned, $Y^\ast=1+1=2$.
\end{enumerate}
\end{example}

The Gaussian kernel also makes the role of kernel sensitivity transparent. For the scope $\{U_X,U_Y\}$, the relative variances determine which admissible coordinate tends to carry the contrast: $\xi=\sigma_Y^2/(\sigma_Y^2+\sigma_X^2)$ shifts the posterior mean of $X^\ast$ between the local-noise and ancestor explanations. Thus the scope answers which explanations are admissible, while $(\sigma_X^2,\sigma_Y^2)$ express how strongly the analyst prefers one type of admissible revision over another. Reporting results over a range of these parameters is a direct sensitivity analysis when no single cross-world kernel is scientifically privileged.

\begin{table}[t]
\centering
\small
\caption{Comparison of intervention and backtracking scopes for the contrast $Y^\ast=3$ in \eqref{eq:linear-scm}. The scope determines which background coordinates may account for the contrast; the kernel determines the plausibility of alternatives within that scope.}
\label{tab:example-comparison}
\begin{tabular}{>{\raggedright\arraybackslash}p{0.18\textwidth}>{\raggedright\arraybackslash}p{0.42\textwidth}>{\raggedright\arraybackslash}p{0.27\textwidth}}
\toprule
Semantics/Scope & How the contrast is realized & Consequence \\
\midrule
Hard $\doo(Y=3)$ & mechanism for $Y$ is replaced & $(X^\ast,Z^\ast)=(1,3)$ \\
$\mathcal E=\{U_Y\}$ & local noise of $Y$ changes & $(1,3)$ \\
$\mathcal E=\{U_X\}$ & upstream $X$ changes & $(2,4)$ \\
$\mathcal E=\{U_X,U_Y\}$ & upstream and local noises share the contrast & distribution over $(X^\ast,Z^\ast)$ with $U_Z$ pinned \\
Full $\mathcal E=\U$ & every background coordinate may change & adds uncertainty from $U_Z^\ast$, not needed for $Y^\ast=3$ \\
$\mathcal E=\{U_Z\}$ & only downstream noise may change & antecedent infeasible \\
\bottomrule
\end{tabular}
\end{table}

\section{Application Case: Diagnostic Explanations and Recourse}
\label{sec:adult-application}

We illustrate the framework on the UCI Adult benchmark \citep{adultUCI}, a standard tabular dataset for predicting whether annual income exceeds \$50K from demographic and socioeconomic variables. The dataset has also been used in causal fairness and recourse \citep{vonKugelgen2022fairness}. We adopt the feature graph studied by \citet{plecko2020fair}, shown in Figure~\ref{fig:adult-dag}, and append a learned binary classifier
\begin{equation*}
Y=f_\theta(X),
\qquad
X=(E,R,J,O),
\end{equation*}
where $E$ is education, $R$ workclass, $J$ hours per week, and $O$ occupation. Non-root structural assignments are fitted with random forest regressors or classifiers according to variable type, and $f_\theta$ is itself a random-forest classifier; full preprocessing, learner, structural-fitting, scope, and kernel details are given in Appendix~\ref{app:adult-application-details}. The fitted decision achieves accuracy $0.789$ and ROC-AUC $0.797$ on the held-out test set.

For individuals receiving the unfavorable decision $Y=0$, we consider the diagnostic explanation query in \eqref{eq:query} with desired outcome $y^\ast=1$ and under different scopes $\mathcal E$ and a Gaussian-copula sensitivity kernel indexed by $\rho\in[0,1]$. Within the scoped coordinates, $\rho=0$ corresponds to independent factual and counterfactual draws, whereas $\rho=1$ makes them identical. We consider $\rho\in\{0,0.5,0.9\}$ and focus on
\begin{equation*}
\mathcal E_{\mathrm{edu}}=\{U_E\},
\qquad
\mathcal E_X=\{U_E,U_R,U_J,U_O\},
\qquad
\mathcal E_{\mathrm{full}}=\U.
\end{equation*}
The first permits revision only of the education disturbance, the second permits revision of disturbances local to the explanatory vector $X$, and the third allows full backtracking.

The resulting explanations exhibit the structural distinction motivating scope restriction. Under $\mathcal E_{\mathrm{edu}}$ and $\rho=0$, the equal-case average of the case-specific conditional probabilities $\mathbb P(E^\ast>E\mid Y^\ast=1,H=h)$ is approximately $0.996$; the corresponding averages for increasing hours and changing occupation are $0.766$ and $0.633$. Thus, although $U_J$ and $U_O$ are pinned, $J^\ast$ and $O^\ast$ need not remain factual: they can change through propagation of the revised education value along the original structural equations. Scope therefore restricts which background disturbances may carry the contrast, rather than directly fixing downstream endogenous variables.

Scope determines which coordinates may be revised, while the kernel determines how probability mass is distributed across the permitted revisions. Averaged over 200 factual negative cases, at $\rho=0$, the mean case-specific values of $\mathbb P(Y^\ast=1\mid H=h)$ are $0.076$, $0.082$, and $0.089$ under $\mathcal E_{\mathrm{edu}}$, $\mathcal E_X$, and $\mathcal E_{\mathrm{full}}$, respectively. Thus, in this fitted benchmark, broader scopes assign more probability to counterfactual backgrounds that reach the favorable decision. Increasing $\rho$ has the opposite effect: at $\rho=0.9$, the corresponding means fall to $0.032$, $0.041$, and $0.051$. Since larger $\rho$ forces the counterfactual state to be closer to the factual one, successful revisions become less likely. 

The application also gives a direct instance of Proposition~\ref{prop:local-reducibility-mtp}. Since $Y=f_\theta(X)$, taking the endogenous frontier to be $A=X$ intercepts every directed route to $Y$. Moreover, since $f_\theta$ is a fixed deterministic function once trained, setting $X=\tilde x$ fixes the target pair $(X,Y)$ to $(\tilde x,f_\theta(\tilde x))$, so query--response compatibility holds automatically. Hence the scoped diagnostic explanation admits a modified-value policy representation on observable $X$. Because $H=(X,Y,Z)$, $A^\natural=X$, and $Y=f_\theta(X)$, the $Y$-component of $H$ is redundant once $(Z,X^\natural)$ is known. Thus, in this application, the induced policy can be indexed only by pre-frontier $Z$ and the natural frontier value $X^\natural$, giving the representation a  modified-value policy interpretation. Interpreting it specifically as recourse additionally requires the components of $X$ to be regarded as jointly actionable.

We use the fitted model to illustrate the marginal bounds derived in Appendix~\ref{app:partial-identification}. Let
\begin{equation*}
F=\{Y=0\},
\qquad
T^\ast=\{Y^\ast=1\},
\qquad
I^\ast=\{E^\ast\geq13,Y^\ast=1\}.
\end{equation*}
The target
\begin{equation*}
\mathbb P(E^\ast\geq13
\given
Y=0,Y^\ast=1,S)
\end{equation*}
asks, within each sex stratum ($S$), among individuals with an unfavorable factual decision who attain a favorable counterfactual decision, how often the successful revision has $E^\ast\geq13$, corresponding to a bachelor's degree or above on the Adult education-number scale.

If the fully specified fitted cross-world law is replaced by only its factual and counterfactual margins, the correspondence between factual negatives and counterfactual successes is lost. As a result, the margins alone are largely uninformative for the conditional query: in $16$ of the $18$ scope--kernel--sex configurations, they permit zero probability for the conditioning event, giving the trivial envelope $[0,1]$. Providing just one extra cross-world quantity already sharpens the result considerably. If the overlap $\eta=\mathbb P(F\cap T^\ast\given S)$ is known, thereby fixing the amount of probability mass shared by factual negatives and counterfactual successes, the median interval width across configurations falls to $0.122$. In every case, the refined interval still contains the corresponding value under the fully specified fitted law.

\section{Discussion and Conclusion}\label{sec:discussion}

We introduced \emph{scope-restricted backtracking counterfactuals}, which refine full backtracking by allowing only a designated subset of exogenous coordinates to differ across factual and counterfactual worlds. The framework separates the \emph{scope}, determining where a contrast may arise, from the cross-world \emph{kernel}, determining which alternatives are plausible, while revised exogenous values still propagate to endogenous descendants. We also characterize when restricted-backtracking queries admit modified-value policy representations and derive partial-identification bounds under limited cross-world information.

The framework nevertheless relies on modeling assumptions that should be made explicit. Latent exogenous variables can be handled through posterior abduction, but the cross-world kernel is not identified from observational data and must be scientifically specified or varied in sensitivity analysis. Corollary~\ref{cor:graphical-frontier} provides sufficient rather than necessary conditions, and the sharp bounds in Appendix~\ref{app:partial-identification} apply to event-valued rather than pointwise density queries. Finally, the Adult analysis uses a benchmark SCM rather than an identified graph, and its Monte Carlo summaries do not propagate learner uncertainty.


\section*{Code availability}
Code implementing the examples and reproducing the associated results is available at \url{https://github.com/johandh2o/scope-restricted-backtracking-counterfactuals}.

\section*{Acknowledgments}
This work was supported by the Research Council of Norway, Integreat --- Norwegian Centre for knowledge-driven machine learning, project number 332645.

\bibliography{references}

\appendix

\section{Preliminaries, Notation, and Theoretical Background}\label{app:preliminaries}

\subsection{Structural causal models and notation}\label{sec:structural}

A \emph{structural causal model} (SCM) is a tuple $\M=(\V,\U,\Gm,\mathcal F,P_{\U})$, where $\V$ is a finite set of endogenous variables, $\U$ is a finite set of background noise variables, $\Gm$ is a directed acyclic graph (DAG) over $\V$, $\mathcal F=\{f_V\}_{V\in\V}$ is a collection of measurable structural assignments, and $P_{\U}$ is a probability measure on $\dom \U$ \citep{PearlCausality}. For each $V\in\V$, let $U_V$ denote the subvector of exogenous coordinates entering the structural assignment for $V$. Thus
\begin{equation}\label{eq:SEM}
    V=f_V\bigl(\pa(V;\Gm),U_V\bigr)\quad P_{\U}\text{--almost surely}.
\end{equation}
In the Markovian case, the subvectors $U_V$ are node-specific and mutually independent. In the semi-Markovian case, latent common exogenous variables are represented graphically by bidirected edges and the projected graph over observables is an acyclic directed mixed graph (ADMG) \citep{evans2014markovian}. 

We assume the structural assignments admit a unique solution for every $u\in\dom\U$, as is guaranteed in particular by acyclicity. Let $\Upsilon_{\M}:\dom\U\to \dom\V$ denote the measurable solution map obtained by recursively solving \eqref{eq:SEM} along a topological order of $\Gm$.

For $A\subseteq\V$, $\pa(A;\Gm)$, $\an(A;\Gm)$, $\de(A;\Gm)$, and $\nde(A;\Gm)$ denote its parents, ancestors, descendants, and non-descendants.

\subsection{Hard and soft intervention and $\sigma$-regimes}\label{sec:soft}

For $A\in\V$, $a\in\dom A$, and $u\in\dom\U$, the hard-intervention counterfactual $Y^a(u)$ is the value of the $Y$-component evaluated at $u$ in the recursive solution map $\Upsilon_{\M}^a$ of the structural equations obtained by replacing the assignment for $A$ by the constant assignment $A=a$. The intervened graph removes arrows entering $A$ in $\Gm$ \citep{PearlCausality} and the corresponding population-level interventional law is
\begin{equation*}
\mathbb{P}(Y\in\mathcal{Y} \given \doo(A=a))
\equiv \mathbb{P}(Y^a\in\mathcal{Y}) :=
\rmint\! \mathbb{I}\{Y^a(u)\in\mathcal{Y}\}\,\dd P_{\U}(u),
\qquad \mathcal{Y}\subseteq\dom Y,
\end{equation*}
with density or mass $p_Y(y \given \doo(A=a))$. Hard interventions provide the canonical semantics for action-oriented questions but are less natural for diagnostic questions that ask how the background conditions would have had to differ for an event to occur on-manifold.

Several formulations of stochastic and soft interventions have been proposed, including modified treatment policies \citep{haneuse2013estimation,diaz2023nonparametric}, shift intervention policies \citep{sani2020identification}, incremental propensity-score interventions (IPI) \citep{kennedy2018nonparametric}, cost-interpolated IPIs \citep{deAguas_2026}, and generalized policies \citep{levis2024stochastic}. One general SCM formulation is given by \citet{Correa_Bareinboim_2020,NeuripsCorrea} through $\sigma$-regimes, which we adopt and present in a convenient unit-level form. 

Fix $A\in\V$ and a unit with exogenous realization $\U=u$. Let $\omega$ be an auxiliary noise variable on a measurable space $\Omega$ with conditional law $P_{\omega\given\U}(\cdot\given u)$. A (unit-level) \emph{soft intervention} or \emph{$\sigma$-regime} on $A$ is a mechanism replacement
\begin{equation*}
\mathfrak{i}_A(u,\omega)=(X,g,\omega),
\end{equation*}
where $X\subseteq \nde(A;\Gm)$ is a replacement parent set (within the original non-descendants of $A$), and $g:\dom X\times \dom U_A\times \Omega\to \dom A$ is a measurable replacement mechanism. The intervention replaces the original assignment of $A$ at $u$
\begin{equation*}
    A(u)=f_A(\pa(A;\Gm)(u),u_A) \quad\text{by}\quad A(u,\omega)=g(X(u),u_A,\omega),
\end{equation*}
leaving all other assignments untouched. Let $\Upsilon ^{\mathfrak{i}_A}_\mathcal{M}$ denote the solution map of the new structural equations and $Y^{\mathfrak{i}_A}(u,\omega)$ be its $Y$-component at $(u,\omega)$. Then one can denote 
\begin{equation*}
\mathbb{P}(Y^{\mathfrak{i}_A}\in\mathcal{Y})
:=
\rmint\!\!\!\rmint\! \mathbb{I}\!\left\{Y^{\mathfrak{i}_A}(u,\omega)\in\mathcal{Y}\right\}\,
\dd P_{\omega\given\U}(\omega\given u)\,\dd P_{\U}(u).
\end{equation*}
For marginal queries, this can also be denoted with the semi-colon notation of \citet{Correa_Bareinboim_2020,NeuripsCorrea}: $\mathbb{P}(Y\in\mathcal{Y} \fatsemi \mathfrak{i}_A)$. When $\omega$ is absent, we work with the singleton space $\{\varnothing\}$ and the degenerate law $\dd P_{\omega\given\U}=\delta_{\varnothing}$. The associated \emph{operated graph} $\Gm^{\mathfrak{i}_A}$ deletes incoming arrows into $A$, adds arrows $X\to A$, and may mark the regime node $\mathfrak{i}_A\to A$ \citep{dawid2002}.

The standard intervention classes are recovered as special cases. Hard interventions correspond to $X=\varnothing$ with constant-valued $g$ and degenerate $\dd P_{\omega\given\U}=\delta_{\varnothing}$ (noise-free). Conditional interventions $\doo\{A=h(X)\}$ correspond to noise-free $g(X,u_A,\varnothing)=h(X)$. Pure stochastic interventions reproducing a conditional distribution $P^\dagger(\cdot\given X)$ correspond to $\omega\sim\mathrm{Unif}[0,1]$, with $\omega\indep\U$, and the conditional-quantile transform $g(X,u_A,\omega)=\inf\{a:P^\dagger(a\given X)\ge\omega\}$. Except in the atomic case, the regime-level law $p_Y(y\fatsemi\mathfrak{i}_A)$ is in general distinct from the conditional law $p_Y(y\given A=a\fatsemi\mathfrak{i}_A)$ \citep{Correa_Bareinboim_2020,NeuripsCorrea}.

\section{Derivations and Proofs}

\subsection{Simultaneous $\sigma$-representation of restricted backtracking}
\label{app:sigma-representations}

For completeness, restricted backtracking can be written equivalently as a simultaneous $\sigma$-regime. By a simultaneous $\sigma$-regime we mean a collection $\{\mathfrak{i}_V:V\in\V\}$ of mechanism replacements applied jointly and driven by one shared auxiliary variable $\omega$, whose conditional law may depend on the factual background $\U=u$. The resulting acyclic system is solved using the same draw $\omega$ in every replacement; here we take $\omega=u^\ast=(e^\ast,\bar e)$, where $e^\ast\sim K(\cdot\given e,\bar e)$. Given a factual background $u=(e,\bar e)$ and such a scoped revision $e^\ast$, define
\begin{equation*}
\mathfrak{i}(u,e^\ast)
=
\{\mathfrak{i}_V(u,e^\ast):V\in\V\},
\qquad
\mathfrak{i}_V(u,e^\ast)
=
\bigl(\pa(V;\Gm),g_V,u^\ast\bigr),
\end{equation*}
with
\begin{equation*}
g_V(x,u_V,u^\ast):=f_V(x,u_V^\ast),
\end{equation*}
where $f_V$ is the original causal mechanism for $V$ in the SCM. Thus every structural assignment retains its original parent set and
structural map, but is evaluated using the same partially revised
background $u^\ast$.

Since $\Gm$ is acyclic, solving the resulting assignments in topological order is exactly the ordinary SCM solution at $u^\ast$. Hence, pointwise,
\begin{equation*}
Y^{\mathfrak{i}}(u,e^\ast)
=
Y(u^\ast)
=
Y(e^\ast,\bar e),
\end{equation*}
while factual variables remain evaluated at $u=(e,\bar e)$. Integrating
over $e^\ast\sim K(\cdot\given e,\bar e)$ and
$u\sim P_{\U}$ therefore gives
\begin{equation*}
\mathbb P(Y^{\mathfrak{i}}\in\mathcal Y,Z\in\mathcal Z)
=
\mathbb P(Y^\ast\in\mathcal Y,Z\in\mathcal Z),
\end{equation*}
which is precisely the scoped cross-world law in
\eqref{eq:partialback}. Thus the simultaneous $\sigma$-regime is an
equivalent notation for restricted backtracking.

\subsection{Proof of Proposition~\ref{prop:support-monotonicity}}
\label{app:support-monotonicity}

Write $\mathcal D=\mathcal E_2\setminus\mathcal E_1$, and let $d(u)$ denote the factual value of the coordinates in $\mathcal D$. Fix a factual background $u$. Under scope $\mathcal E_1$, any feasible counterfactual background has the form
\begin{equation*}
u^\ast=(e_1^\ast,d(u),\bar e_2), \quad\text{where}\quad e_1^\ast\in\operatorname{supp}\{K_{\mathcal E_1}(\cdot\given u)\},
\end{equation*}
and $\bar e_2$ denotes the factual value of the coordinates outside $\mathcal E_2$. Hence
\begin{equation*}
(e_1^\ast,d(u))
\in
\operatorname{supp}\{K_{\mathcal E_1}(\cdot\given u)\}\times\{d(u)\}.
\end{equation*}

By assumption,
\begin{equation*}
\operatorname{supp}\{K_{\mathcal E_1}(\cdot\given u)\}\times\{d(u)\}
\subseteq
\operatorname{supp}\{K_{\mathcal E_2}(\cdot\given u)\},
\end{equation*}
so $(e_1^\ast,d(u))\in\operatorname{supp}\{K_{\mathcal E_2}(\cdot\given u)\}$. Since the coordinates outside $\mathcal E_2$ are fixed at the same factual values under both scopes, the same full counterfactual background
\begin{equation*}
u^\ast=(e_1^\ast,d(u),\bar e_2)
\end{equation*}
is therefore feasible under $\mathcal E_2$ whenever it is feasible under $\mathcal E_1$.

Consequently, any antecedent $v=V(u^\ast)$ attainable under $\mathcal E_1$ is also attainable under $\mathcal E_2$. Hence every antecedent feasible under $\mathcal E_1$ is feasible under $\mathcal E_2$. The result does not impose an ordering on incomparable scopes. For example, Example~\ref{ex:linear} gives $\{U_Z\}$ and $\{U_Y\}$ as an incomparable pair. \hfill$\square$

\subsection{Proof of Proposition~\ref{prop:local-reducibility-mtp} and Corollary~\ref{cor:graphical-frontier}}
\label{app:local-reducibility-proof}

\paragraph{Proof of Proposition~\ref{prop:local-reducibility-mtp}.}
Fix $H=h$ and write
\begin{equation*}
Q:=(W,A^\natural).
\end{equation*}
Under the scoped backtracking construction, condition~\eqref{eq:functional-frontier} implies
\begin{equation}\label{eq:proof-frontier-identity}
C^\ast
=
C^{A^\ast}(u)
\qquad
\text{almost surely given }H=h.
\end{equation}

Let $\varphi$ be any bounded measurable function on the state space of $C=(X,Y)$. Then, using~\eqref{eq:proof-frontier-identity},
\begin{align}
\E\!\left[
\varphi(C^\ast)
\given H=h
\right]
&=
\rmint\!
\E\!\left[
\varphi\{C^a(u)\}
\given
A^\ast=a,Q=q,H=h
\right]
\,\dd P_{A^\ast,Q}(a,q\given H=h).
\label{eq:proof-disintegration-1}
\end{align}
Query--response compatibility~\eqref{eq:query--response-compatibility} removes the additional conditioning on the selected frontier value from the response term. Since
\begin{equation*}
\nu_h(\cdot\given q)
=
P_{A^\ast}(\cdot\given Q=q,H=h),
\end{equation*}
equation~\eqref{eq:proof-disintegration-1} becomes
\begin{align}
\E\!\left[
\varphi(C^\ast)
\given H=h
\right]
&=
\rmint\!\!\!\rmint\!
\E\!\left[
\varphi\{C^a(u)\}
\given Q=q,H=h
\right]
\dd \nu_h(a\given q)\,
\dd P_Q(q\given H=h).
\label{eq:proof-backtracking-mixture}
\end{align}

Now consider the modified-value policy. Conditional on $(Q,H=h)$, the policy draws
\begin{equation*}
A^{\nu_h}\sim\nu_h(\cdot\given Q)
\end{equation*}
independently of the remaining factual latent background and propagates the original structural assignments. Hence
\begin{equation*}
C^{\nu_h}
=
C^{A^{\nu_h}}(u),
\end{equation*}
and the conditional independence built into the policy yields
\begin{align}
\E\!\left[
\varphi(C^{\nu_h})
\given H=h
\right]
&=
\rmint\!\!\!\rmint\!
\E\!\left[
\varphi\{C^a(u)\}
\given Q=q,H=h
\right]
\dd \nu_h(a\given q)\,
\dd P_Q(q\given H=h).
\label{eq:proof-policy-mixture}
\end{align}
The right-hand sides of~\eqref{eq:proof-backtracking-mixture} and~\eqref{eq:proof-policy-mixture} coincide for every bounded measurable $\varphi$. Therefore
\begin{equation*}
C^\ast\given (H=h)
\quad\text{and}\quad
C^{\nu_h}\given (H=h)
\end{equation*}
have the same probability law. Since $C=(X,Y)$,
\begin{equation}\label{eq:proof-joint-law}
(X^\ast,Y^\ast)\given (H=h)
\overset{d}{=}
(X^{\nu_h},Y^{\nu_h})\given (H=h).
\end{equation}
Whenever the corresponding conditional density or mass function exists and the conditioning value has positive normalizing mass, equality of the joint laws in~\eqref{eq:proof-joint-law} gives
\begin{equation*}
p_{X^\ast}(x^\ast\given Y^\ast=y^\ast,H=h)
=
p_{X^{\nu_h}}
(x^\ast\given Y^{\nu_h}=y^\ast,H=h),
\end{equation*}
which is~\eqref{eq:query--reducibility}.
\hfill$\square$

\paragraph{Proof of Corollary~\ref{cor:graphical-frontier}(i).}
Fix a factual background $u=(e,\bar e)$ compatible with $H=h$, an admissible scoped revision $e^\ast$, and let $u^\ast=(e^\ast,\bar e)$ and $a^\ast=A(u^\ast)$. Compare (a) the ordinary SCM solution under $u^\ast$ with (b) the solution under factual background $u$ after $\operatorname{do}(A=a^\ast)$. By construction, the two solutions agree at every node in $A$.

If a target $V^{\dagger}\in X\cup Y$ belongs to $A$, the two solutions therefore agree at $V^{\dagger}$. Otherwise, condition (i) gives $V^{\dagger}\notin S_{\mathcal E}$, so a discrepancy at $V^{\dagger}$ cannot originate from a changed local exogenous argument at $V^{\dagger}$.

Let $\Delta\subseteq\V$ denote the set of nodes at which the two solutions differ. As just noted, $\Delta\cap A=\varnothing$. Moreover, every $V'\in\Delta$ either satisfies $U_{V'}\cap\mathcal E\ne\varnothing$, so that $V'\in S_{\mathcal E}\setminus A$, or else has at least one parent in $\Delta$, since otherwise both its parent values and its local exogenous argument would coincide across the two constructions and its assignment $f_{V'}$ would return the same value. Because $\Gm$ is acyclic, induction along a topological order therefore shows that every $V'\in\Delta$ is reached from some $V\in S_{\mathcal E}\setminus A$ by a directed path lying entirely within $\Delta$, and hence disjoint from $A$. If some target $V^{\dagger}\in(X\cup Y)\setminus A$ belonged to $\Delta$, this would produce a directed path $V\rightarrow\!\cdot\!\rightarrow V^{\dagger}$ avoiding $A$, contradicting~\eqref{eq:graphical-frontier}.

Hence the two constructions agree at both target nodes,
\begin{equation*}
C(u^\ast)=C^{A(u^\ast)}(u),
\end{equation*}
which is the downstream-frontier condition~\eqref{eq:functional-frontier}.\hfill$\square$

\paragraph{Proof of Corollary~\ref{cor:graphical-frontier}(ii).}
Fix $\tilde a$ and write $Q=(W,A^\natural)$ and $L=(Q,H)$. Conditional on the factual latent state $\U$, $C^{\tilde a}$ is fixed by the SCM, whereas the randomness in $A^\ast$ is supplied by the backtracking kernel. The first two conditional independences in~\eqref{eq:structural-comp} therefore imply
\begin{equation*}
P_{A^\ast}(a^\ast\given \U,L)
=
P_{A^\ast}(a^\ast\given U_{\mathrm{sel}},L),
\qquad
P_{C^{\tilde a}}(c\given \U,L)
=
P_{C^{\tilde a}}(c\given U_{\mathrm{res}},L).
\end{equation*}
Because $C^{\tilde a}$ is a deterministic function of $\U$, the pair $(A^\ast,C^{\tilde a})$ is conditionally independent given $(\U,L)$, so iterated expectation gives
\begin{equation*}
P_{A^\ast,C^{\tilde a}}(a^\ast,c\given L) =
\mathbb E\!\left[
P_{A^\ast}(a^\ast\given U_{\mathrm{sel}},L)
P_{C^{\tilde a}}(c\given U_{\mathrm{res}},L)
\given L
\right].
\end{equation*}
Since $U_{\mathrm{sel}}\indep U_{\mathrm{res}}\given L$, this factorizes as
\begin{equation*}
P_{A^\ast,C^{\tilde a}}(a^\ast,c\given L)
=
P_{A^\ast}(a^\ast\given L)\,
P_{C^{\tilde a}}(c\given L),
\end{equation*}
and hence
\begin{equation*}
A^\ast\indep C^{\tilde a}\given L.
\end{equation*}
Evaluating at $A^\ast=\tilde a$ gives
\begin{equation*}
P_{C^{\tilde a}}
\bigl(c\given A^\ast=\tilde a,W,A^\natural,H\bigr)
=
P_{C^{\tilde a}}
\bigl(c\given W,A^\natural,H\bigr),
\end{equation*}
which is query--response compatibility~\eqref{eq:query--response-compatibility}.
\hfill$\square$

\subsection{Understanding query--response compatibility}
\label{app:frontier-compatibility}

The downstream-frontier condition and query--response compatibility address different requirements. The former is structural: once the revised frontier value $A^\ast$ is known, it asks whether the upstream scoped revision can be replaced by $\operatorname{do}(A=A^\ast)$. Query--response compatibility is instead probabilistic: after this replacement, it asks whether independently redrawing the selected frontier value would discard latent information that matters for the corresponding response.

To see this, write $Q=(W,A^\natural)$ and $L=(Q,H)$. Compatibility requires
\begin{equation*}
P_{C^a}
\bigl(
c\given A^\ast=a,L
\bigr)
=
P_{C^a}
\bigl(
c\given L
\bigr)
\end{equation*}
for every $c$. Thus, $A^\ast=a$ may reveal latent information beyond $L$; the requirement is only that this information be irrelevant to the response $C^a$ under the value actually selected. In particular, it does \emph{not} require the stronger condition
\begin{equation}\label{eq:full-latent-independence}
A^\ast
\indep \U
\given W,A^\natural,H.
\end{equation}
The following two examples illustrate the distinction.

\paragraph{Example 1: harmless latent dependence.}
Consider the Markovian SCM
\begin{equation*}
Z=f_Z(U_Z),
\qquad
X=\mathbb I\{U_X\leq\pi(Z)\},
\qquad
Y=g(X,Z,U_Y),
\end{equation*}
with mutually independent $(U_Z,U_X,U_Y)$. Restrict backtracking to
$\mathcal E=\{U_X\}$ and let
\begin{equation*}
U_X^\ast\given(U_X=u_X)
\sim
\mathcal{N}(u_X,\sigma^2).
\end{equation*}
Take the frontier to be $A=X$ and let $W=Z$.

Conditioning on the factual value of $X$ reveals only which side of the threshold contains $U_X$. Since the backtracking kernel is centered at the factual $U_X$, the selected value $X^\ast$ generally retains additional
information about $U_X$, so
\begin{equation*}
X^\ast\not\!\indep U_X\given X,Z
\end{equation*}
in general. Hence the stronger latent-independence condition~\eqref{eq:full-latent-independence} need not hold.

This residual dependence is nevertheless harmless for the response. Under the intervention $X=a$,
\begin{equation*}
C^a
=
(X^a,Y^a)
=
\bigl(a,g(a,Z,U_Y)\bigr),
\end{equation*}
which no longer depends on $U_X$. In the terminology of Corollary~\ref{cor:graphical-frontier}(ii), one may take $U_{\mathrm{sel}}=U_X$ and $U_{\mathrm{res}}=U_Y$: the selected frontier value depends on the former, the response depends on the latter, and their conditional separation follows from the Markovian exogenous structure. Hence the additional information about $U_X$ conveyed by $X^\ast=a$ does not change the conditional distribution of $C^a$. Query--response compatibility can therefore hold even though $X^\ast$ remains associated with the factual latent background.

\paragraph{Example 2: response-relevant latent selection.}
Now consider a factual exogenous background $(U_X,U_R)$ satisfying
\begin{equation*}
\mathbb P(U_X=0,U_R=0)
=
\mathbb P(U_X=1,U_R=1)
=
\frac12,
\end{equation*}
and define
\begin{equation*}
X=\mathbb I\{U_X=2\},
\qquad
Y=XU_R.
\end{equation*}
Under the factual distribution, $(X,Y)=(0,0)$ almost surely, so factual evidence $H=(X,Y)=(0,0)$ does not distinguish the two latent backgrounds.

Restrict the scope to $\mathcal E=\{U_X\}$ and set $U_X^\ast=U_X+1$. With frontier $A=X$,
\begin{equation*}
X^\ast
=
\mathbb I\{U_X^\ast=2\}
=
\mathbb I\{U_X=1\}
=
U_R.
\end{equation*}
Moreover, the downstream-frontier condition holds: once $X^\ast$ is fixed, the scoped change can be reproduced by intervening on $X$ at that value.

Compatibility nevertheless fails because the selected frontier value now reveals a latent response modifier. Under $X=1$,
\begin{equation*}
Y^1=U_R,
\end{equation*}
and since $X^\ast=U_R$,
\begin{equation*}
    \mathbb P
\bigl(
Y^1=1\given X^\ast=1,H
\bigr)
=
1,
\qquad
\mathbb P
\bigl(
Y^1=1\given H
\bigr)
=
\frac12.
\end{equation*}
Here $U_{\mathrm{sel}}=U_X$ and $U_{\mathrm{res}}=U_R$ are perfectly dependent, so the latent-separation requirement in Corollary~\ref{cor:graphical-frontier}(ii) fails. Accordingly, $X^\ast=1$ selects precisely those factual backgrounds with the high response $Y^1=1$. An independently redrawn modified-value policy can match the conditional distribution of $X^\ast$, but it cannot preserve this coupling between frontier selection and the latent response modifier.

The two examples isolate the role of query--response compatibility. Dependence between the backtracked frontier value and the factual latent background is not itself an obstacle to reducibility. Reducibility fails only when the dependence discarded by independently redrawing the frontier value carries response information under that same selected value.

\section{Partial Identification}\label{app:partial-identification}

This section develops partial-identification results for the scoped explanation query in \eqref{eq:query} when the joint dependence between factual and counterfactual variables is only partially specified. The sharp marginal result is stated for discrete or event-valued explanations and depends on selected factual and counterfactual margins rather than on observational information alone; for continuous variables, the corresponding statement concerns measurable events of positive conditional probability, not pointwise densities. Additional denominator information yields explicit refinements, while kernel restrictions can be incorporated through a feasible-set refinement when they can be translated into constraints on the relevant cross-world overlaps.

\subsection{Sharp marginal bounds}\label{app:bounds}

We first consider the information contained in selected margins alone. Because \eqref{eq:query} conditions on factual information $Z=z$, let
\begin{equation*}
\mathbb P_z(\cdot):=\mathbb P(\cdot\given Z=z),
\end{equation*}
with $\mathbb P_z=\mathbb P$ when no additional $Z$ is present. We first write the construction for discrete $X$ and $Y$ and then state the event-valued analogue, which also covers measurable-set queries for continuous variables.

For a candidate explanation $x^\ast$, define
\begin{equation*}
F:=\{X=x,Y=y\},\qquad
T^\ast:=\{Y^\ast=y^\ast\},\qquad
I^\ast:=\{X^\ast=x^\ast,Y^\ast=y^\ast\}.
\end{equation*}
For example, suppose $Y$ indicates whether a credit application is approved and $X$ indicates whether the applicant has high debt. Then $F=\{X=1,Y=0\}$ represents applicants who are factually high-debt and rejected, $T^\ast=\{Y^\ast=1\}$ represents counterfactual worlds in which the application is approved, and $I^\ast=\{X^\ast=0,Y^\ast=1\}$ represents counterfactual worlds in which the applicant has low debt and is approved. Thus $I^\ast$ represents one specific candidate explanation for the desired outcome encoded by $T^\ast$. We have $I^\ast\subseteq T^\ast$ and
\begin{equation*}
\mathbb P_z(I^\ast\given F\cap T^\ast)
=
\mathbb P(X^\ast=x^\ast\given Y^\ast=y^\ast,H=h).
\end{equation*}
Suppose the three stratum-specific margins
\begin{equation*}
\phi_z:=\mathbb P_z(F),\qquad
\tau_z:=\mathbb P_z(T^\ast),\qquad
\gamma_z:=\mathbb P_z(I^\ast)
\end{equation*}
are identified or otherwise specified, and define
\begin{align*}
r_z&:=\max\{0,\phi_z+\tau_z-1\},
&
s_z&:=\min\{\phi_z,\tau_z\},\\
\ell_z&:=\max\{0,\phi_z+\gamma_z-1\},
&
m_z&:=\min\{\phi_z,\gamma_z\}.
\end{align*}
\begin{proposition}[Sharp marginal explanation bounds]
\label{prop:xai-pearl-bounds}
Under the setup above, suppose $r_z>0$, so that $F\cap T^\ast$ has positive probability under every cross-world coupling compatible with the specified margins. For a discrete candidate $x^\ast$, the sharp marginal bounds are
\begin{equation}
\label{eq:xai-pearl-bounds}
p_{\mathrm{low},z}(x^\ast)
=
\frac{\ell_z}{r_z},
\qquad
p_{\mathrm{upp},z}(x^\ast)
=
\min\left\{1,\frac{m_z}{r_z}\right\}.
\end{equation}
Both endpoints are attainable over all cross-world couplings compatible with $(\phi_z,\tau_z,\gamma_z)$. More generally, if the point events defining $F$, $T^\ast$, and $I^\ast$ are replaced by measurable events of positive conditional probability, the same formulas sharply bound $\mathbb P_z(I^\ast\given F\cap T^\ast)$. Thus, for continuous variables the result applies to event-valued explanations; it does not by itself provide pointwise density or MAP-density bounds.
\end{proposition}

For discrete candidate explanations, a candidate $x^\ast$ cannot be a MAP explanation when
\begin{equation*}
p_{\mathrm{upp},z}(x^\ast)
<
\sup_{\tilde x}p_{\mathrm{low},z}(\tilde x),
\end{equation*}
and it is the unique MAP explanation whenever
\begin{equation*}
p_{\mathrm{low},z}(x^\ast)
>
\sup_{\tilde x\neq x^\ast}p_{\mathrm{upp},z}(\tilde x).
\end{equation*}

\paragraph{Proof and sharpness.}
Let
\begin{equation*}
a_z:=\mathbb P_z(F\cap I^\ast),
\qquad
b_z:=\mathbb P_z(F\cap T^\ast),
\end{equation*}
so that the target equals $a_z/b_z$. Fr\'echet inequalities and $I^\ast\subseteq T^\ast$ imply
\begin{equation*}
\ell_z\leq a_z\leq m_z,\qquad
r_z\leq b_z\leq s_z,\qquad
0\leq a_z\leq b_z,\qquad
0\leq b_z-a_z\leq\tau_z-\gamma_z.
\end{equation*}
For a fixed feasible denominator $b$, the smallest and largest feasible numerators are therefore
\begin{equation*}
a_{\min}(b)
=
\max\{\ell_z,b-(\tau_z-\gamma_z)\},
\qquad
a_{\max}(b)
=
\min\{m_z,b\}.
\end{equation*}
Hence the smallest and largest attainable ratios at $b$ are
\begin{equation*}
\max\left\{
\frac{\ell_z}{b},
1-\frac{\tau_z-\gamma_z}{b}
\right\},
\qquad
\min\left\{
1,\frac{m_z}{b}
\right\},
\end{equation*}
respectively. Write $k_z:=\tau_z-\gamma_z$, so that $a_{\min}(b)/b$ equals $\ell_z/b$ for $b\leq\ell_z+k_z$ and $1-k_z/b$ thereafter; the first branch is decreasing and the second increasing, so this map is not monotone on all of $(0,\infty)$. Since $\ell_z=\max\{0,r_z-k_z\}$, however, one has $\ell_z+k_z=\max\{r_z,k_z\}$. If $r_z\geq k_z$ this gives $\ell_z+k_z=r_z$, so only the increasing branch is active on $[r_z,s_z]$; if $r_z<k_z$ then $\ell_z=0$ and the decreasing branch contributes the constant value $0$. In either case $b\mapsto a_{\min}(b)/b$ is nondecreasing on $[r_z,s_z]$, so the lower ratio is minimized at $b=r_z$. Since $b\mapsto\min\{m_z,b\}/b$ is nonincreasing, the upper ratio is also maximized at $b=r_z$. This yields \eqref{eq:xai-pearl-bounds}.

Both endpoints are attainable by allocating the factual mass $F$ across the three disjoint events $I^\ast$, $T^\ast\setminus I^\ast$, and $\lnot T^\ast$, whose probabilities
are $\gamma_z$, $\tau_z-\gamma_z$, and $1-\tau_z$. Hence the bounds are sharp.\hfill$\square$

Intuitively, the bounds quantify how much factual mass can be assigned to the candidate explanation $I^\ast$ among all cross-world couplings consistent with the supplied margins. Here $r_z$ is the smallest forced overlap between the factual event and the desired counterfactual outcome, while $\ell_z$ and $m_z$ are the smallest and largest possible overlaps with the candidate explanation. Thus the bounds tighten when the margins force more candidate-specific overlap or limit how much mass $I^\ast$ can contain. The condition $r_z>0$ guarantees a nonvanishing query denominator.

Notice that $\phi_z$ is observational, whereas $\tau_z$ and $\gamma_z$ are counterfactual margins and must come from structural assumptions, a specified SCM and backtracking kernel, external information, a valid modified-value reduction, or sensitivity ranges.

\subsection{Refinements}
\label{app:query--refined-bounds}

The marginal bounds deliberately allow every cross-world coupling compatible with the three specified margins. Thus, any additional information that rules out some of these couplings can sharpen the same query. Let
\begin{equation*}
a_z:=\mathbb P_z(F\cap I^\ast),
\qquad
\eta_z:=\mathbb P_z(F\cap T^\ast),
\end{equation*}
so that the target is $a_z/\eta_z$, and define the margin-compatible region
\begin{equation*}
\mathcal B_z
:=
\left\{
(a,b):
\ell_z\leq a\leq m_z,\;
r_z\leq b\leq s_z,\;
0\leq a\leq b,\;
0\leq b-a\leq\tau_z-\gamma_z
\right\}.
\end{equation*}
If additional information $\mathcal I$ can be translated into an exact restriction of the admissible pairs to $\mathcal B_z(\mathcal I)\subseteq\mathcal B_z$, then the corresponding sharp refinement is
\begin{equation}
\label{eq:kernel-aware-query-bounds}
\inf_{(a,b)\in\mathcal B_z(\mathcal I):\,b>0}\frac{a}{b}
\leq
\mathbb P(X^\ast=x^\ast\given Y^\ast=y^\ast,H=h)
\leq
\sup_{(a,b)\in\mathcal B_z(\mathcal I):\,b>0}\frac{a}{b}.
\end{equation}
Thus extra information can shrink the feasible numerator--denominator region, and the explanation probability is re-optimized over the smaller set. When $\mathcal B_z(\mathcal I)$ is only a conservative outer restriction rather than an exact characterization, the same optimization yields valid but not necessarily sharp bounds. Importantly, joint restrictions should be imposed directly on $(a,b)$ rather than converted into separate numerator and denominator bounds, which can unnecessarily lose sharpness.

\paragraph{Kernel restrictions.}
Information about the scope $\mathcal E$ and backtracking kernel $K$ can rule out cross-world couplings that are allowed by the margins alone. If these restrictions can be translated into a set
$\mathcal B_{\mathcal E,K}(z)\subseteq\mathcal B_z$ of compatible numerator--denominator pairs, then
\eqref{eq:kernel-aware-query-bounds} applies with
$\mathcal I=(\mathcal E,K)$; it is sharp when that feasible set is characterized exactly. Intuitively, the kernel describes which counterfactual backgrounds are plausible given the factual one, and therefore constrains which numerator--denominator combinations can actually arise. The display therefore provides a general refinement template rather than a closed-form kernel-specific bound: a concrete kernel must first be translated into restrictions on $(a,b)$. Such information may be available when the backtracking kernel is substantively specified as part of the model, calibrated from domain knowledge, or restricted through scientifically meaningful features such as locality or similarity. 

\paragraph{Known denominator.}
A particularly useful intermediate case arises when the numerator remains unknown but
\begin{equation*}
\eta_z=\mathbb P_z(F\cap T^\ast)
\end{equation*}
is known. In words, $\eta_z$ is the probability, within the factual stratum
$Z=z$, that the factual event $F$ holds and the desired counterfactual outcome $T^\ast$ is realized for the same unit. For instance, if $Y$ indicates whether a credit application is approved and $X$ indicates whether the applicant has high debt, then for $F=\{X=1,Y=0\}$ and $T^\ast=\{Y^\ast=1\}$, $\eta_z$ is the probability, among applicants with $Z=z$, of being factually high-debt and rejected while being approved in the counterfactual world.

If $ 0<\eta_z\in[r_z,s_z]$ then the feasible numerator is exactly restricted to
\begin{equation*}
\max\{0,\ell_z,\eta_z-(\tau_z-\gamma_z)\}
\leq a_z\leq
\min\{m_z,\eta_z\}.
\end{equation*}
Provided this interval is nonempty, the resulting sharp bounds are
\begin{equation}
\label{eq:query-denominator-refined}
p^{\eta}_{\mathrm{low},z}(x^\ast)
=
\frac{\max\{0,\ell_z,\eta_z-(\tau_z-\gamma_z)\}}{\eta_z},
\qquad
p^{\eta}_{\mathrm{upp},z}(x^\ast)
=
\frac{\min\{m_z,\eta_z\}}{\eta_z}.
\end{equation}
Knowing $\eta_z$ removes uncertainty about the total factual--counterfactual overlap entering the query denominator; only its allocation between $I^\ast$ and $T^\ast\setminus I^\ast$ remains unknown.

An exactly known denominator is stronger information than the marginal inputs and will not usually be directly available from observational data. Alternatively $\eta_z$ can itself be treated as a sensitivity parameter whose variation measures how strongly the explanation depends on the factual--counterfactual overlap.

\paragraph{Bounded denominator.}
Suppose instead that one knows or assumes
\begin{equation*}
\eta_z^-\leq\eta_z\leq\eta_z^+.
\end{equation*}
If $\mathcal{I}_z:=[\eta_z^-,\eta_z^+]\cap[r_z,s_z]$ is nonempty and bounded away from zero, then
\begin{equation*}
\inf_{b\in \mathcal{I}_z}
\frac{\max\{0,\ell_z,b-(\tau_z-\gamma_z)\}}{b}
\leq
\mathbb P(X^\ast=x^\ast\given Y^\ast=y^\ast,H=h)
\leq
\sup_{b\in \mathcal{I}_z}
\frac{\min\{m_z,b\}}{b}.
\end{equation*}
By the same monotonicity argument as in Appendix~\ref{app:bounds}, the lower ratio is nondecreasing and the upper ratio is nonincreasing over the feasible denominator range, so both extrema are attained at the smallest feasible $b\in \mathcal{I}_z$.

This case is often more realistic than specifying $\eta_z$ exactly: scientific knowledge may only constrain the fraction of factual units that could reach $Y^\ast=y^\ast$, or external information may provide an interval rather than a point value. The endpoints $(\eta_z^-,\eta_z^+)$ therefore have a direct sensitivity-analysis interpretation and can be widened or narrowed to assess robustness.

\subsection{Toy Bound Analysis}\label{app:bound-analysis}

This subsection illustrates the partial-identification hierarchy with simple simulated cross-world laws using the credit example above. Let $Y$ indicate whether a credit application is approved, $X$ whether the applicant has high
debt, and $Z$ additional observed applicant information, such as an income category. For simplicity, we omit $Z$ here, so that $\mathbb P_z=\mathbb P$.
We take
\begin{equation*}
F=\{X=1,Y=0\}
\end{equation*}
to represent applicants who are factually high-debt and rejected,
\begin{equation*}
T^\ast=\{Y^\ast=1\}
\end{equation*}
to represent counterfactual approval, and
\begin{equation*}
I_j^\ast=\{X^\ast=j,Y^\ast=1\},\qquad j\in\{0,1\},
\end{equation*}
to represent approval accompanied by counterfactual debt status $j$. Thus $\phi=\mathbb P(F)$, $\tau=\mathbb P(T^\ast)$, and $\gamma_j=\mathbb P(I_j^\ast)$. The examples do not estimate a full SCM; they only illustrate what the bounds imply once these margins are supplied. The same calculations apply within any fixed stratum $Z=z$.

\paragraph{A uniquely ranked MAP explanation.}
First, simulate $n=10^6$ units with
\begin{equation*}
\mathbb{I}(F)\sim \operatorname{Bernoulli}(0.7),
\qquad
\mathbb{I}(T^\ast)\sim \operatorname{Bernoulli}(0.6),
\end{equation*} 
independently, and, among units satisfying $T^\ast$, set $X^\ast=1$ with probability $11/12$ and $X^\ast=0$ otherwise. The resulting estimates are
\begin{equation*}
\widehat\phi=0.700,
\qquad
\widehat\tau=0.601,
\qquad
\widehat\gamma_1=0.551,
\qquad
\widehat\gamma_0=0.050.
\end{equation*}
Applying \eqref{eq:xai-pearl-bounds} gives
\begin{equation*}
p_{\mathrm{low}}(1)=0.834,
\qquad
p_{\mathrm{upp}}(1)=1,
\end{equation*}
whereas
\begin{equation*}
p_{\mathrm{low}}(0)=0,
\qquad
p_{\mathrm{upp}}(0)=0.166.
\end{equation*}
Hence
\begin{equation*}
p_{\mathrm{low}}(1)>p_{\mathrm{upp}}(0),
\end{equation*}
so $X^\ast=1$ is the unique MAP explanation for every cross-world coupling compatible with the supplied margins. In the credit example, among applicants
who are factually high-debt and rejected but would be approved in the
counterfactual world, the most likely accompanying debt state is necessarily
still high debt. The corresponding simulated probabilities are approximately
$0.917$ for $X^\ast=1$ and $0.083$ for $X^\ast=0$, although these values are
not used to construct the bounds.

\paragraph{An inconclusive marginal analysis.}
Now simulate $n=10^6$ units with
\begin{equation*}
\mathbb{I}(F)\sim \operatorname{Bernoulli}(0.65),
\qquad
\mathbb{I}(T^\ast)\sim \operatorname{Bernoulli}(0.5),
\end{equation*}
independently, and, among units satisfying $T^\ast$, set $X^\ast=1$ with probability $1/2$ and $X^\ast=0$ otherwise. The estimated margins are
\begin{equation*}
\widehat\phi=0.650,
\qquad
\widehat\tau=0.499,
\qquad
\widehat\gamma_1=0.250,
\qquad
\widehat\gamma_0=0.250.
\end{equation*}
For either debt state, \eqref{eq:xai-pearl-bounds} gives
\begin{equation*}
p_{\mathrm{low}}(j)=0,
\qquad
p_{\mathrm{upp}}(j)=1,
\qquad j\in\{0,1\}.
\end{equation*}
Thus the margins alone cannot determine whether counterfactual approval is more likely to accompany high or low debt among the factually rejected applicants. In the simulated law both conditional probabilities are close to $0.5$, but many other cross-world couplings share the same margins.

\paragraph{Refinement using denominator information.}
Finally, simulate $n=10^6$ units with
\begin{equation*}
\mathbb{I}(T^\ast)\sim \operatorname{Bernoulli}(0.5),
\qquad
\mathbb P\{\mathbb{I}(F)=1\given \mathbb{I}(T^\ast)=1\}=0.7,
\qquad
\mathbb P\{\mathbb{I}(F)=1\given \mathbb{I}(T^\ast)=0\}=0.6,
\end{equation*}
and, among units satisfying $T^\ast$, set $X^\ast=1$ with probability $0.6$. The estimated margins are
\begin{equation*}
\widehat\phi=0.649,
\qquad
\widehat\tau=0.499,
\qquad
\widehat\gamma_1=0.300.
\end{equation*}
Using only these margins, \eqref{eq:xai-pearl-bounds} gives the vacuous interval
\begin{equation*}
0\leq
\mathbb P(I_1^\ast\given F\cap T^\ast)
\leq 1.
\end{equation*}
Thus the margins alone provide no information about the probability that a factually high-debt, rejected applicant who would instead be approved would also remain high-debt in that counterfactual world.

Suppose, however, that additional information determines
\begin{equation*}
\eta=\mathbb P(F\cap T^\ast),
\end{equation*} 
whose population value under the simulated cross-world law is $0.35$ and whose Monte Carlo estimate is $\widehat\eta=0.349$. Here $\eta$ is the probability of being factually high-debt and rejected while being approved in the counterfactual world. The known-denominator refinement in
\eqref{eq:query-denominator-refined} yields
\begin{equation*}
0.430
\leq
\mathbb P(I_1^\ast\given F\cap T^\ast)
\leq
0.860.
\end{equation*}
Indeed,
\begin{equation*}
p_{\mathrm{low}}^{\eta}(1)
=
\frac{
\max\{0,\widehat\phi+\widehat\gamma_1-1,
\widehat\eta-(\widehat\tau-\widehat\gamma_1)\}
}{
\widehat\eta
}
=0.430,
\end{equation*}
whereas
\begin{equation*}
p_{\mathrm{upp}}^{\eta}(1)
=
\frac{
\min\{\widehat\phi,\widehat\gamma_1,\widehat\eta\}
}{
\widehat\eta
}
=0.860.
\end{equation*}
The true simulated value is approximately $0.6$. Thus the explanation is only partially identified: among applicants who are factually high-debt and rejected but would be approved counterfactually, the probability that counterfactual approval is accompanied by remaining high-debt lies between roughly $43\%$ and $86\%$. Hence high debt remains a plausible, and potentially dominant, counterfactual explanation state among approval worlds, but because its identified interval crosses $0.5$, the available information does not determine whether high or low debt is the more likely explanation state.

\section{Details for the Adult Benchmark Application}
\label{app:adult-application-details}

This appendix gives the structural specification, fitted learners and backtracking mechanism, and empirical analyses underlying Section~\ref{sec:adult-application}. We focus on three questions: how scope changes a diagnostic explanation, when that explanation reduces exactly to a modified-value policy on observable features, and how much information is lost when the cross-world coupling is not specified.

\subsection{Benchmark and structural specification}

We use the UCI Adult benchmark \citep{adultUCI}, which contains 48,842 census-derived records with demographic and socioeconomic variables and is commonly used to predict whether annual income exceeds \$50K. Adult has also been widely used in the causal fairness and recourse literature, including by \citet{vonKugelgen2022fairness}. Its limitations as a substantive benchmark for fairness, however, are well documented \citep{ding2021retiring}. Accordingly, we use Adult here only for comparability with prior work and as a convenient fitted benchmark SCM for illustrating scope-restricted queries; we draw no substantive policy or fairness conclusions about income from the analysis.

We adopt the feature graph used in the Adult analysis of \citet{plecko2020fair}, whose causal structure has previously been studied and operationalized, and append a learned decision node $Y$; the resulting graph is displayed in Figure~\ref{fig:adult-dag}. After removing observations with missing values, the analysis contains 45,222 complete records, split into 31,655 training and 13,567 test observations. We define
\begin{equation*}
Y=f_\theta(X),
\qquad
X=(E,R,J,O)
=
(\text{education},\text{workclass},\text{hours/week},\text{occupation}),
\end{equation*}
where $f_\theta$ is a random-forest classifier trained to predict the observed income label. Its test accuracy and ROC-AUC are $0.789$ and $0.797$, respectively; the observed positive rate in the test sample is $0.248$, while the fitted classifier predicts a positive decision for $0.113$ of test observations. The graph is used as a benchmark structural specification rather than as a claim that the Adult observational distribution identifies the underlying causal structure.

Table~\ref{tab:adult-learners} lists the learners and distributional representations used in the fitted benchmark.
\begin{table}[t]
\centering
\caption{Learners used in the fitted Adult benchmark.}
\label{tab:adult-learners}
\small
\begin{tabular}{p{0.36\linewidth}p{0.57\linewidth}}
\toprule
Component & Learner / representation \\
\midrule
Numerical SCM node with parents
& Random-forest regressor + empirical residual quantiles \\
Categorical SCM node with parents
& Random-forest classifier + inverse-CDF latent ranks \\
Numerical SCM root
& Empirical marginal distribution \\
Categorical SCM root
& Empirical category probabilities \\
Decision $Y=f_\theta(X)$
& Random-forest classifier \\
\bottomrule
\end{tabular}
\end{table}
All random forests use 200 trees and minimum leaf size 20. For $f_\theta$, numerical components of $X$ enter directly and categorical components are one-hot encoded; the binary decision is obtained by thresholding the fitted probability at $0.5$.

\begin{figure*}[t]
\centering

\tikzset{
  var/.style={inner sep=1pt, font=\small},
  exo/.style={inner sep=1pt, font=\small},
  int/.style={draw, rectangle, rounded corners=1pt, inner sep=2pt, font=\small},
  arr/.style={->, >=Stealth, line width=1.0pt},
  back/.style={<->, >=Stealth, densely dashed, line width=1.0pt}
}

\begin{minipage}[c]{0.48\textwidth}
\centering
\begin{tikzpicture}[scale=0.7]

\node[var] (D) at (0, 1.8) {$D$};
\node[var] (M) at (0, 0)   {$M$};
\node[int] (E) at (0,-1.8) {$E$};

\node[int] (R) at (4.0, 1.8) {$R$};
\node[int] (J) at (4.0, 0)   {$J$};
\node[int] (O) at (4.0,-1.8) {$O$};

\node[var] (Y) at (6.3,0) {$Y$};


\draw[arr] (D) -- (M);
\draw[arr] (M) -- (E);

\draw[arr] (D.west)
    to[out=180,in=180,looseness=1.15]
    (E.west);


\draw[arr] (D) -- (R);
\draw[arr] (M) -- (J);
\draw[arr] (E) -- (O);

\draw[arr] (D) -- (J);
\draw[arr] (M) -- (R);

\draw[arr] (M) -- (O);
\draw[arr] (E) -- (J);

\draw[arr] (D.north)
    to[out=25,in=155,looseness=1.30]
    (O.north);

\draw[arr] (E.south)
    to[out=-25,in=-155,looseness=1.30]
    (R.south);


\draw[arr] (R) -- (Y);
\draw[arr] (J) -- (Y);
\draw[arr] (O) -- (Y);

\draw[arr] (E.south)
    to[out=-65,in=-120,looseness=1.20]
    (Y.south);

\end{tikzpicture}

\vspace{-1cm}
{\small (a) Adult benchmark feature DAG}
\end{minipage}
\hfill
\begin{minipage}[c]{0.48\textwidth}
\centering
\vspace{1.5cm}
\begin{tikzpicture}[scale=0.7]

\node[var] (D) at (0,0) {$D$};
\node[var] (M) at (1.55,0) {$M$};
\node[int] (X) at (3.75,0) {$X$};
\node[var] (Y) at (6.25,0) {$Y=f_\theta(X)$};

\node[exo,red] (UM) at (1.35,1.2) {$U_M$};
\node[exo,red] (UX) at (4.05,1.2)
{$(U_E,U_R,U_J,U_O)$};

\draw[arr,gray] (UM) -- (M);
\draw[arr,gray] (UX) -- (X);

\draw[arr] (D) -- (M);
\draw[arr] (D) to[bend left=-25] (X);
\draw[arr] (M) -- (X);
\draw[arr] (X) -- (Y);

\end{tikzpicture}
\vspace{1.5cm}\\
{\small (b) Scoped revision and frontier $A=X$}
\end{minipage}

\caption{
Adult benchmark graph used in the application. Here $S$ is sex, $G$ age, $N$ native country, $M$ marital status, $E$ education, $R$ workclass, $J$ hours per week, and $O$ occupation; panel~(a) groups the three root variables as $D=(S,G,N)$ to display the literature graph compactly. The explanatory vector is $X=(E,R,J,O)$ and the learned decision is $Y=f_\theta(X)$; boxed nodes indicate the frontier $A=X$. Thus, taking the candidate endogenous frontier in Section~\ref{sec:soft-bridge} to be $A=X$, every directed path from a scoped upstream revision to $Y$ intersects the frontier. Panel~(b) emphasizes that scopes are imposed on exogenous coordinates, while the corresponding modified-value representation acts on the endogenous variable $X$.
}
\label{fig:adult-dag}
\end{figure*}

\paragraph{Fitted structural model and backtracking kernel.}
We fit the structural assignments in the adopted DAG sequentially on the training sample. For a numerical non-root node, a random-forest regressor estimates its conditional mean given its parents, and the fitted structural disturbance is represented through the empirical quantile of the corresponding residual. Numerical roots instead use their empirical marginal distributions. For categorical non-root nodes, a random-forest classifier estimates conditional category probabilities and the structural assignment is represented through the corresponding inverse conditional CDF; categorical roots use empirical marginal category probabilities.

For categorical nodes, the observed value identifies an interval of latent ranks rather than a unique $U_V$. Abduction therefore samples $U_V$ uniformly within the fitted conditional-CDF interval associated with the observed category. A fixed lexicographic ordering of categories is used to define the inverse CDF. These choices are part of the fitted benchmark SCM rather than identification claims about the Adult population. As a numerical audit, abduction followed by regeneration reconstructed all eight modeled variables exactly on 1,500 held-out observations; for the numerical nodes, both the mean and maximum absolute reconstruction errors were zero.

For each scoped coordinate $U_V\in\mathcal E$, we use the Gaussian-copula sensitivity kernel
\begin{equation*}
\zeta_V=\Phi^{-1}(U_V),
\qquad
\zeta_V^\ast
=
\rho \zeta_V+\sqrt{1-\rho^2}\,\epsilon_V,
\qquad
U_V^\ast=\Phi(\zeta_V^\ast),
\end{equation*}
where $\epsilon_V\stackrel{\mathrm{iid}}{\sim}\mathcal{N}(0,1)$ independently across scoped coordinates and independently of the abducted factual ranks; unscoped coordinates satisfy $U_V^\ast=U_V$. Thus $\rho=0$ gives an independent marginal redraw of the scoped disturbance, whereas $\rho\rightarrow1$ approaches the factual latent background. We consider $\rho\in\{0,0.5,0.9\}$. The same abducted ranks and Gaussian innovations are reused across scope--$\rho$ conditions to improve paired comparability.

\paragraph{Task 1: Scope-sensitive diagnostic explanations.}
Let $D=(S,G,N)$ denote the grouped demographic root variables, let $Z=(D,M)=(S,G,N,M)$ denote the remaining pre-$X$ factual variables, and let $H=(X,Y,Z)$ denote the factual evidence. For individuals receiving the unfavorable prediction $Y=0$, we study
\begin{equation}
p_{X^\ast}\bigl(
x^\ast
\given
Y^\ast=1,H=h
\bigr)
\quad\text{under }(\mathcal E,\rho),
\label{eq:adult-diagnostic-query}
\end{equation}
using Monte Carlo abduction, scoped revision, and structural propagation. The final analysis uses 200 factual negative cases and 1,000 revised backgrounds per case. We analyze
\begin{equation*}
\mathcal E_{\mathrm{edu}}=\{U_E\},
\qquad
\mathcal E_X=\{U_E,U_R,U_J,U_O\},
\qquad
\mathcal E_{\mathrm{full}}=\U.
\end{equation*}

For each factual case, $\mathbb P(Y^\ast=1\mid H=h)$ is estimated by the proportion of its 1,000 revised worlds attaining $Y^\ast=1$. The conditional explanation summaries are first computed within each case among successful draws and are then averaged equally over cases with at least one successful draw. Table~\ref{tab:adult-scope-results} reports the resulting macro averages.

\begin{table*}[t]
\centering
\caption{Scope-sensitive diagnostic explanations. Columns 2--4 give the mean probability of a successful revision across 200 factual negative cases. The final columns give equal-case averages of the case-specific explanation laws conditional on $Y^\ast=1$ at $\rho=0$.}
\label{tab:adult-scope-results}
\small
\begin{tabular}{lcccccc}
\toprule
& \multicolumn{3}{c}{Success probability} &
\multicolumn{3}{c}{Conditional on $Y^\ast=1$, $\rho=0$} \\
\cmidrule(lr){2-4}\cmidrule(lr){5-7}
Scope
& $\rho=0$
& $\rho=0.5$
& $\rho=0.9$
& $\mathbb P(E^\ast>E)$
& $\mathbb P(J^\ast>J)$
& $\mathbb P(O^\ast\neq O)$ \\
\midrule
$\mathcal E_{\mathrm{edu}}$
& 0.076 & 0.058 & 0.032 & 0.996 & 0.766 & 0.633 \\
$\mathcal E_X$
& 0.082 & 0.066 & 0.041 & 0.916 & 0.772 & 0.865 \\
$\mathcal E_{\mathrm{full}}$
& 0.089 & 0.076 & 0.051 & 0.915 & 0.797 & 0.877 \\
\bottomrule
\end{tabular}
\end{table*}

In these fitted results, broader scopes assign more probability to the favorable counterfactual at every value of $\rho$, whereas increasing $\rho$ reduces this probability for every scope. More importantly, $\mathcal E_{\mathrm{edu}}$ exhibits the specifically structural behavior motivating scope-restricted backtracking: at $\rho=0$, successful case-specific explanation laws almost always satisfy $E^\ast>E$, but they also increase hours with probability $0.766$ and change occupation with probability $0.633$, despite $U_J$ and $U_O$ being pinned. Restricting which exogenous disturbances may change therefore does not freeze downstream endogenous variables.

The high-similarity regime also illustrates that conditional explanation summaries and success probability capture different aspects of the query. At $\rho=0.9$, at least one successful Monte Carlo draw is observed for only $42.5\%$ of factual cases under $\mathcal E_{\mathrm{edu}}$, compared with $66.5\%$ under $\mathcal E_X$ and $80.5\%$ under $\mathcal E_{\mathrm{full}}$. Thus the corresponding conditional summaries at large $\rho$ describe the subset of case-specific laws for which the simulation observes positive success mass. Consistently with the increasing similarity, the mean distance between the factual and revised feature vectors among successful case-specific laws decreases under $\mathcal E_{\mathrm{edu}}$ from $3.07$ at $\rho=0$ to $1.82$ at $\rho=0.9$. Here distance is the mixed-feature sum of standardized absolute differences over the numerical components of $X$ and indicators of disagreement over its categorical components.

\paragraph{Task 2: From diagnostic explanation to recourse.}
The same construction gives a direct application of Proposition~\ref{prop:local-reducibility-mtp}. Since
\begin{equation*}
Y=f_\theta(X),
\qquad
X=(E,R,J,O),
\end{equation*}
taking the endogenous frontier to be $A=X$ intercepts every directed route through which a scoped upstream revision can affect $Y$. Moreover, for $C=(X,Y)$,
\begin{equation*}
C^{\tilde x}
=
(\tilde x,f_\theta(\tilde x)).
\end{equation*}
Although $f_\theta$ is fitted by random forest, once trained it is a fixed deterministic function of $X$. Hence $C^{\tilde x}$ is fixed once $\tilde x$ is specified and query--response compatibility holds automatically. The diagnostic law in \eqref{eq:adult-diagnostic-query} therefore admits the exact modified-value representation
\begin{equation*}
p_{X^\ast}
\bigl(x^\ast\given Y^\ast=1,H=h\bigr)
=
p_{X^{\nu}}
\bigl(x^\ast\given Y^{\nu}=1,H=h\bigr),
\end{equation*}
where, because $H=(X,Y,Z)$, $X^\natural=X$, and the factual decision is deterministic given $X$, the generic policy $\nu_h$ of Proposition~\ref{prop:local-reducibility-mtp} simplifies here to
\begin{equation*}
\nu(\cdot\given z,x^\natural)
=
P_{X^\ast}
\bigl(
\cdot
\given
Z=z,
X^\natural=x^\natural
\bigr).
\end{equation*}
Thus the same explanation generated through revisions of latent exogenous coordinates can be reproduced exactly by a randomized modified-value policy acting on observable $X$. The induced policy uses only pre-frontier $Z$ and the natural frontier value $X^\natural$, giving the representation a prospective randomized-policy interpretation; interpreting it specifically as recourse additionally requires the substantive assumption that the components of $X$ are jointly manipulable or actionable.

\paragraph{Task 3: Information supplied by the cross-world coupling.}
We finally ask how much of the diagnostic explanation remains if the fitted cross-world coupling is discarded. Within each factual-sex stratum, define
\begin{equation*}
F=\{Y=0\},
\qquad
T^\ast=\{Y^\ast=1\},
\qquad
I^\ast=\{E^\ast\geq13,Y^\ast=1\},
\end{equation*}
so the fitted SCM and kernel determine the benchmark query under the fitted model
\begin{equation*}
\mathbb P(E^\ast\geq13
\given
Y=0,Y^\ast=1,S).
\end{equation*}
The partial-identification analysis uses 5,000 factual test observations and 10 revised worlds per observation. Within each factual-sex stratum we retain only
\begin{equation*}
\phi=\mathbb P(F\given S),
\qquad
\tau=\mathbb P(T^\ast\given S),
\qquad
\gamma=\mathbb P(I^\ast\given S),
\end{equation*}
and apply Proposition~\ref{prop:xai-pearl-bounds}.

Across the three primary scopes, three values of $\rho$, and two factual-sex strata, there are 18 configurations. In 16 of them, $r_S=\max\{0,\phi+\tau-1\}=0$, so the margins permit a zero-probability conditioning event, the positivity hypothesis of Proposition~\ref{prop:xai-pearl-bounds} fails, and the marginal-only envelope is the trivial interval $[0,1]$ rather than a sharp bound in the sense of that proposition. The only two non-vacuous marginal-only configurations occur for factual females under full backtracking:
\begin{align*}
\rho=0:
&\qquad
0.925\in[0.679,1],
\\
\rho=0.5:
&\qquad
0.926\in[0.435,1].
\end{align*}
Their widths are $0.321$ and $0.565$, respectively. Providing only the cross-world denominator
\begin{equation*}
\eta=\mathbb P(F\cap T^\ast\given S)
\end{equation*}
substantially sharpens the result. Across all 18 configurations, the median width of the refined interval is $0.122$ and the mean width is $0.177$. The median widths are $0.246$, $0.108$, and $0.104$ under $\mathcal E_{\mathrm{edu}}$, $\mathcal E_X$, and $\mathcal E_{\mathrm{full}}$, respectively. Every refined interval contains the corresponding fully specified fitted-SCM benchmark, and all empirical Fr\'echet and subset-consistency checks are satisfied.

Table~\ref{tab:adult-partial-id} illustrates the result under the $X$-local scope at $\rho=0$. Although the marginal information alone is vacuous for both factual-sex strata, supplying $\eta$ produces narrow intervals around the fitted benchmark.

\begin{table}[t]
\centering
\caption{Partial identification for
$\mathbb P(E^\ast\geq13\mid Y=0,Y^\ast=1,S)$ under
$\mathcal E_X$ and $\rho=0$.}
\label{tab:adult-partial-id}
\small
\begin{tabular}{lccc}
\toprule
Factual sex
& Fitted query
& Marginal only
& Known $\eta$ \\
\midrule
Female & 0.949 & $[0,1]$ & $[0.947,1]$ \\
Male   & 0.913 & $[0,1]$ & $[0.895,1]$ \\
\bottomrule
\end{tabular}
\end{table}

The comparison makes explicit that the separate factual and counterfactual margins generally contain too little information even to guarantee positive mass for the conditioning event in this benchmark. Knowledge of its overlap probability removes this denominator ambiguity and can sharply contract the identified set, while the remaining width captures uncertainty about the unresolved cross-world association.

\end{document}